\documentclass{egpubl}
\usepackage{egsgp17}
\usepackage[british]{babel}

 \JournalSubmission    
 \electronicVersion 

\ifpdf \usepackage[pdftex]{graphicx} \pdfcompresslevel=9
\else \usepackage[dvips]{graphicx} \fi

\PrintedOrElectronic

\usepackage{t1enc,dfadobe}

\usepackage{egweblnk}
\usepackage{cite}
\usepackage{soul}

\usepackage{amssymb}
\usepackage{amsmath}
\usepackage[ruled, vlined, linesnumbered]{algorithm2e}
\usepackage{algorithmic}
\usepackage{booktabs}

\usepackage{soul,color}
\definecolor{black}{rgb}{0,0,0}
\definecolor{turquoise}{cmyk}{0.65,0,0.1,0.1}
\definecolor{purple}{rgb}{0.65,0,0.65}
\definecolor{dark_green}{rgb}{0, 0.5, 0}
\definecolor{blue}{rgb}{0,0,0.8}
\definecolor{brown}{rgb}{0.6,0.5,0.1}
\definecolor{gray}{gray}{0.15}
\definecolor{hgray}{gray}{0.3}
\definecolor{lgray}{gray}{0.5}

\newcommand{\norm}[1]{\left\lVert #1 \right\rVert}

\newcommand{\minorrev}[1]{#1}
\newcommand{\change}[1]{#1}

\title[Robust Structure-based Shape Correspondence]%
      {Robust Structure-based Shape Correspondence}

\author[Yanir Kleiman and Maks Ovsjanikov]{Yanir Kleiman and Maks
  Ovsjanikov\\LIX, Ecole Polytechnique, UMR CNRS}

\begin{document}


\maketitle

\hyphenpenalty=1700

\begin{abstract}
  We present a robust method to find region-level correspondences between shapes, which are invariant
  to changes in geometry and applicable across multiple shape representations.
  We generate simplified shape graphs by jointly decomposing the shapes, and devise an adapted
  graph-matching technique, from which we infer correspondences between shape regions. The
  simplified shape graphs are designed to primarily capture the overall structure of the shapes,
  without reflecting precise information about the geometry of each region, which enables us to find
  correspondences between shapes that might have significant geometric differences. Moreover, due to
  the special care we take to ensure the robustness of each part of our pipeline, our method can
  find correspondences between shapes with different representations, such as triangular meshes and
  point clouds.  We demonstrate that the region-wise matching that we obtain can be used to find
  correspondences between feature points, reveal the intrinsic self-similarities of each shape, and
  even construct point-to-point maps across shapes. Our method is both time and space efficient,
  leading to a pipeline that is significantly faster than comparable approaches. We
  demonstrate the performance of our approach through an extensive quantitative and qualitative
  evaluation on several benchmarks where we achieve comparable or superior performance to existing
  methods.

\begin{classification} 
\CCScat{Computer Graphics}{I.3.5}{Computational Geometry and Object Modeling}{Geometric algorithms}
\end{classification}

\end{abstract}

\section{Introduction}\label{sec:intro}



Finding correspondences between shapes is a fundamental problem in computer graphics and geometry processing. Many
existing methods aim to find point-to-point maps~\cite{kim11,ovsjanikov12}, or correspondences between feature
points~\cite{berg05,leordeanu05,kezurer15}, by minimizing some prescribed distortion energy (e.g. isometric
distortion). Thus, their performance is optimal in case the shapes satisfy the given deformation model, but may fail
under significant distortion, such as non-isometric deformation or missing parts. Several recent approaches have
proposed instead to consider either soft \cite{solomon2012soft}, functional \cite{ovsjanikov12} or region-wise
correspondences \cite{ganapathi16}, which can be especially beneficial under significant deformation, or sampling
changes. However, these approaches typically do not take into account the precise structural properties of the shapes,
and in most cases, do not incorporate global shape connectivity information into the optimization.

To overcome these limitations, we propose a flexible approach that uses global information to
find correspondences between shapes from different classes, which may have very different
geometry. We segment the shapes into regions and build a simplified shape graph, 
which is both robust and applicable to multiple representations. 
To construct the shape graphs, we develop an extension of the \texttt{Mapper} graph
construction~\cite{singh07}, in which we jointly cluster the values of multi-dimensional descriptor functions on the two shapes.
Then, we find a correspondence between regions which
relies on the global structure of the shapes rather than their precise geometry.
\minorrev{Our method is geared towards shapes with similar structure and small topological noise. Thus, we are able to purposely discard some geometric information and rely instead on the structure of the shapes, at the cost of robustness to shapes with strong partiality or significant topological noise.}

\begin{figure}[t]
\centering

	\includegraphics[width=0.6\linewidth]{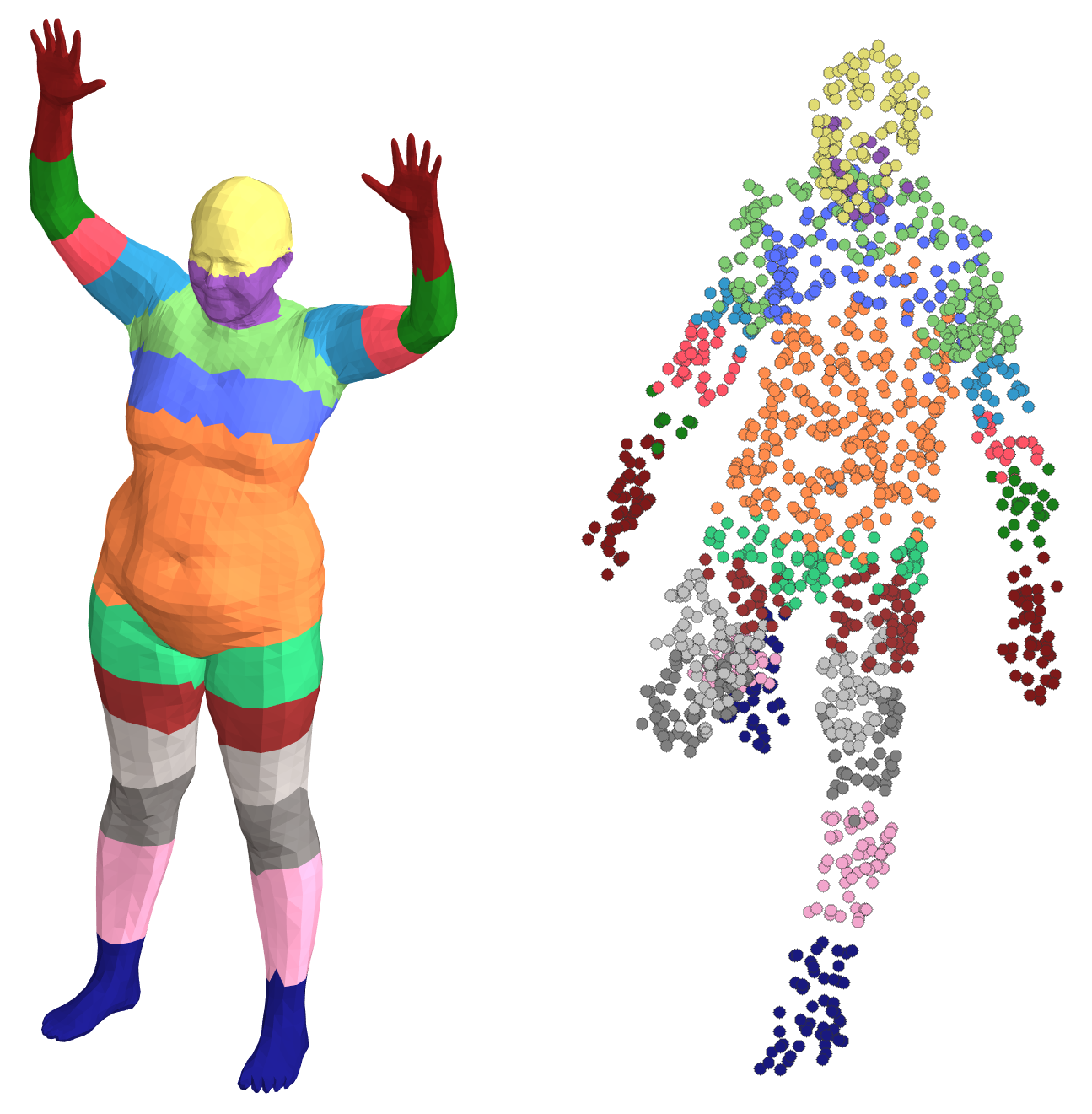}
\caption{\change{Region-based matching between a triangular mesh and a noisy point cloud with a significantly different pose, body type and sampling rate, obtained using our method.}
\vspace{-7mm}
\label{fig:teaser}}

\end{figure}

A correspondence between regions does not provide the same level of details as point-to-point
correspondences. However, we show that region-level correspondences can be used as constraints to
significantly improve point-to-point maps. Often, a point-to-point correspondence is not necessary,
in which case there are numerous advantages to the region-based approach. First, it allows a very
quick optimization compared with existing methods, 
\change{with a core algorithm that is often a magnitude faster. This allows running the process
for several parameter values and selecting the best parameters, leading to a more stable solution
with less distortion between segments compared to state-of-the-art methods.
Finally, we can easily identify parts that \emph{do not have} a match across
shapes.} This allows partial matching of shapes with different topology or missing parts.
Importantly, by working with a significantly simplified representation, we obtain a method that is
both robust and extremely efficient, often requiring orders of magnitude less time and memory
compared to related techniques.

In our solution we pay special attention to matching shapes that have intrinsic symmetries, and as such can pose
problems to many existing methods. \change{We propose a two-step solution that first finds a symmetric correspondence
  between regions, and then breaks the symmetries using a simple and efficient local expansion technique.}  Still, for
  some applications, symmetric matching is enough or might be even more useful than a one-to-one correspondence.  For
  example, an additional application of our method is detection of intrinsic symmetries within a shape by matching the
  shape to itself.
We evaluate our method in a number of experiments, comparing our results with two point-to-point
methods (Blended Intrinsic Maps~\cite{kim11} and functional maps~\cite{ovsjanikov12}), and one
recent region-based method (stable regions~\cite{ganapathi16}), using well-known benchmarks
(TOSCA~\cite{tosca}, SCAPE~\cite{scape} and FAUST~\cite{faust}). 
Additionally, we generate
point-to-point maps from the matching between regions using the functional maps method described
in~\cite{ovsjanikov12}. We show that using the correspondence between regions significantly improves
the accuracy of the map.
Finally, we evaluate our method for matching triangle meshes to point clouds and point clouds to point clouds. We compare these results with~\cite{ganapathi16} for various sampling densities and noise levels, and show a significant improvement in both accuracy and running time.

\begin{figure*}[t]
\centering

	\begin{tabular}{@{}c@{\qquad}c@{\qquad}c@{\qquad}c@{\qquad}c@{}}

		\includegraphics[height=3.5cm]{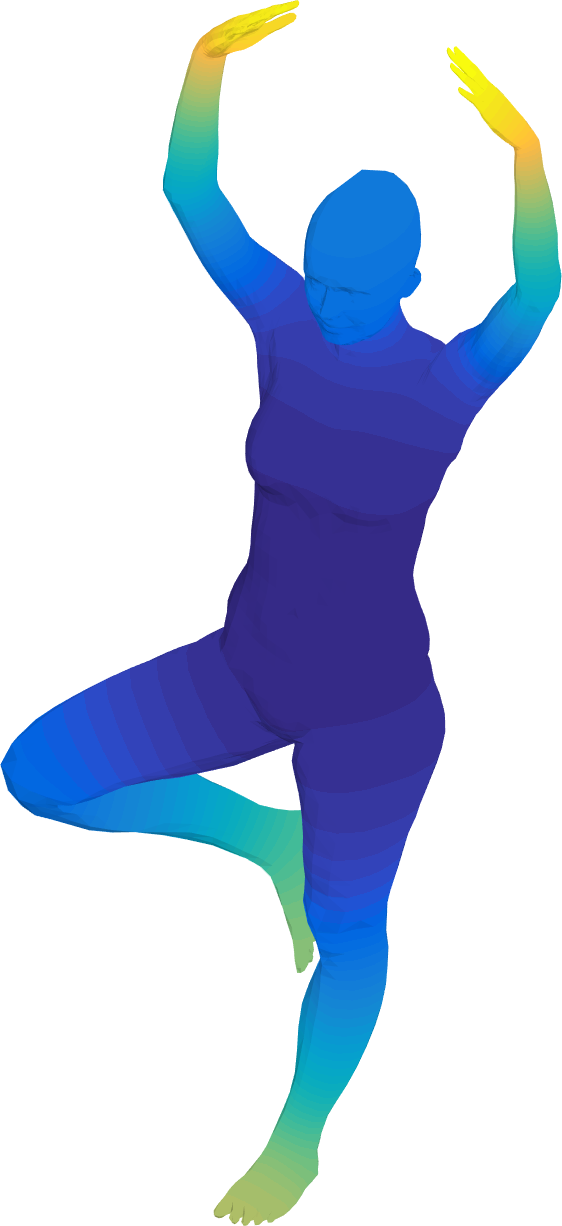} &
		\includegraphics[height=3.5cm]{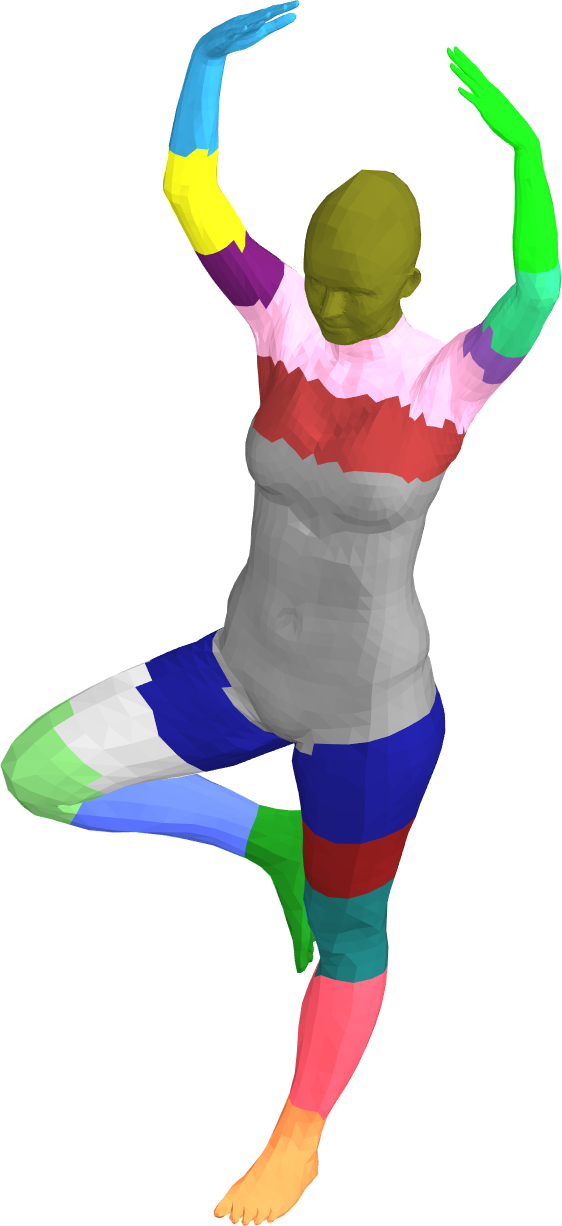} &
		\includegraphics[height=3.5cm]{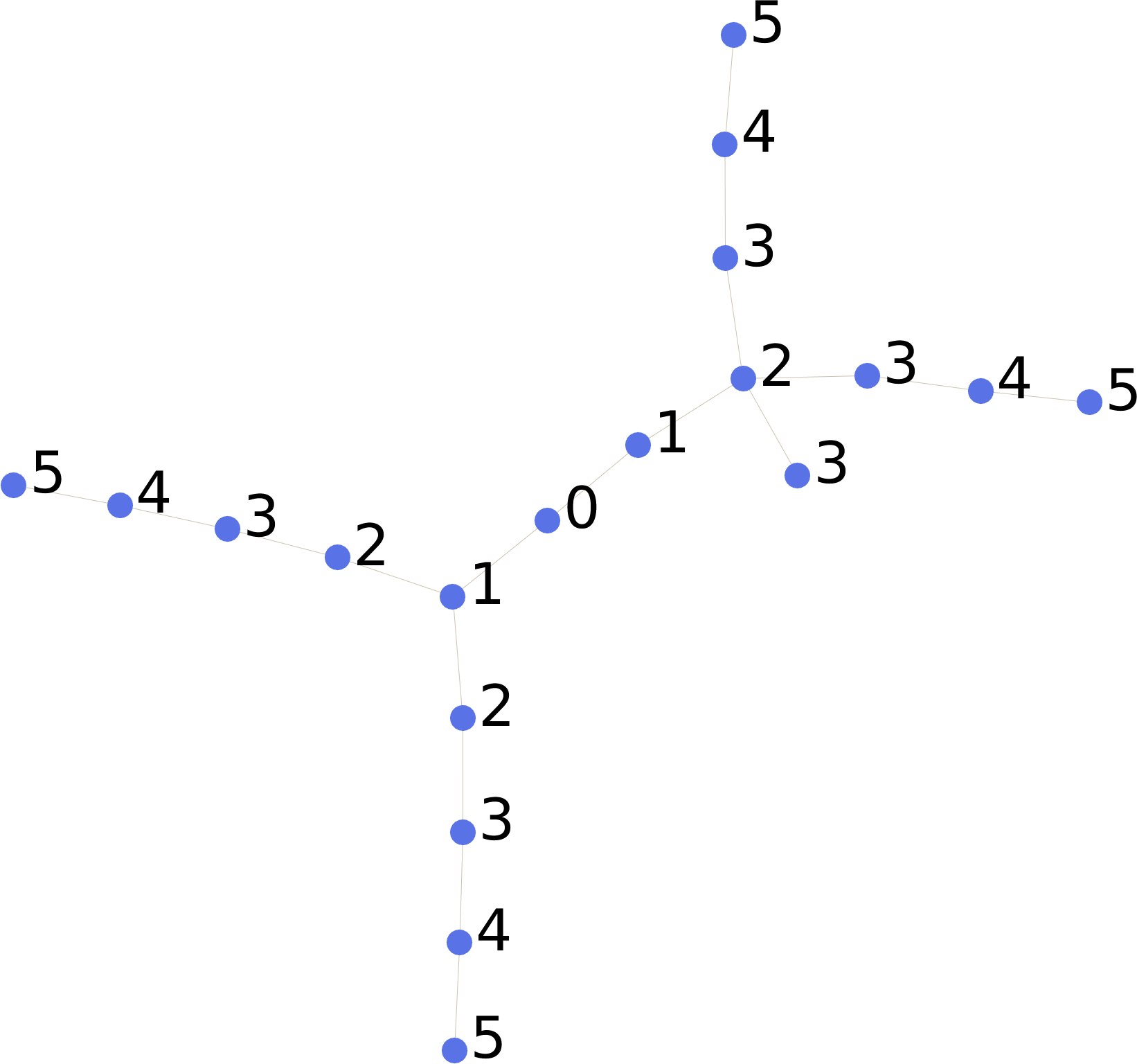} &
		\includegraphics[height=3.5cm]{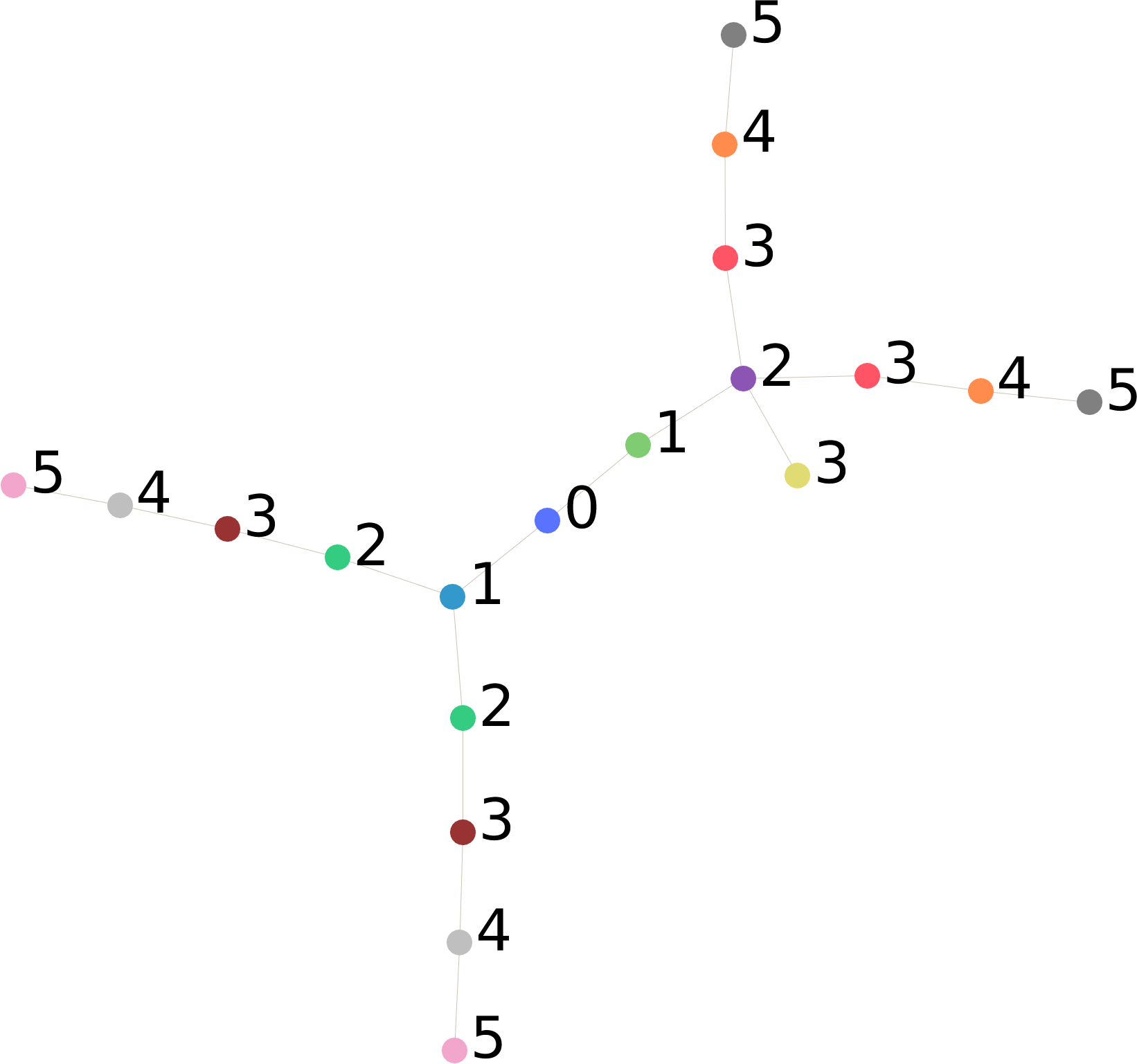} &
		\includegraphics[height=3.5cm]{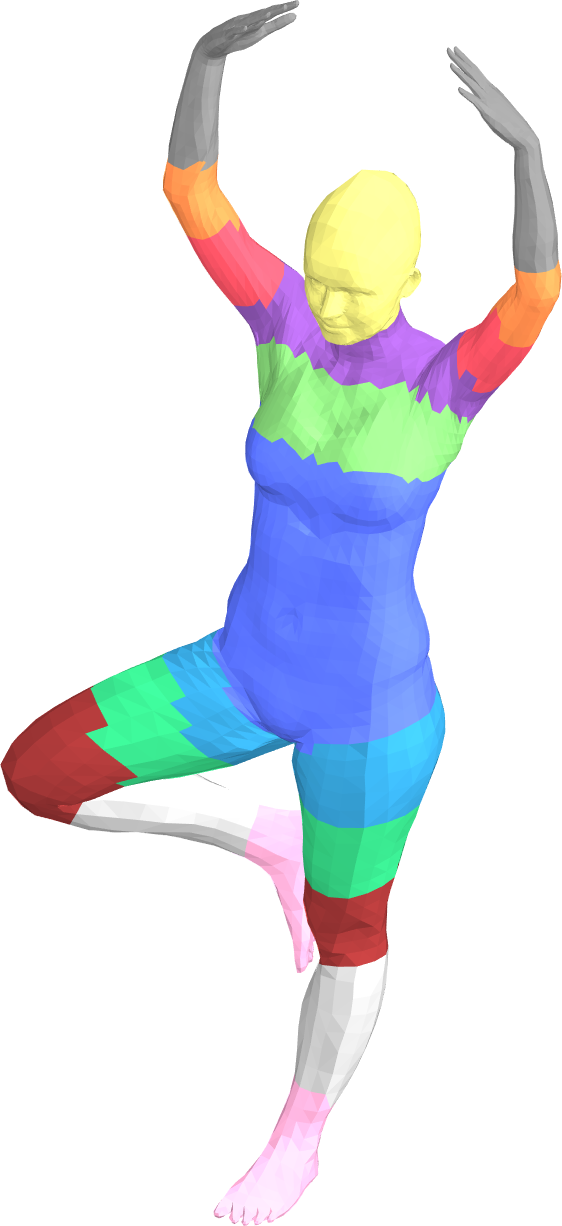} \\
		
		\includegraphics[height=3.5cm]{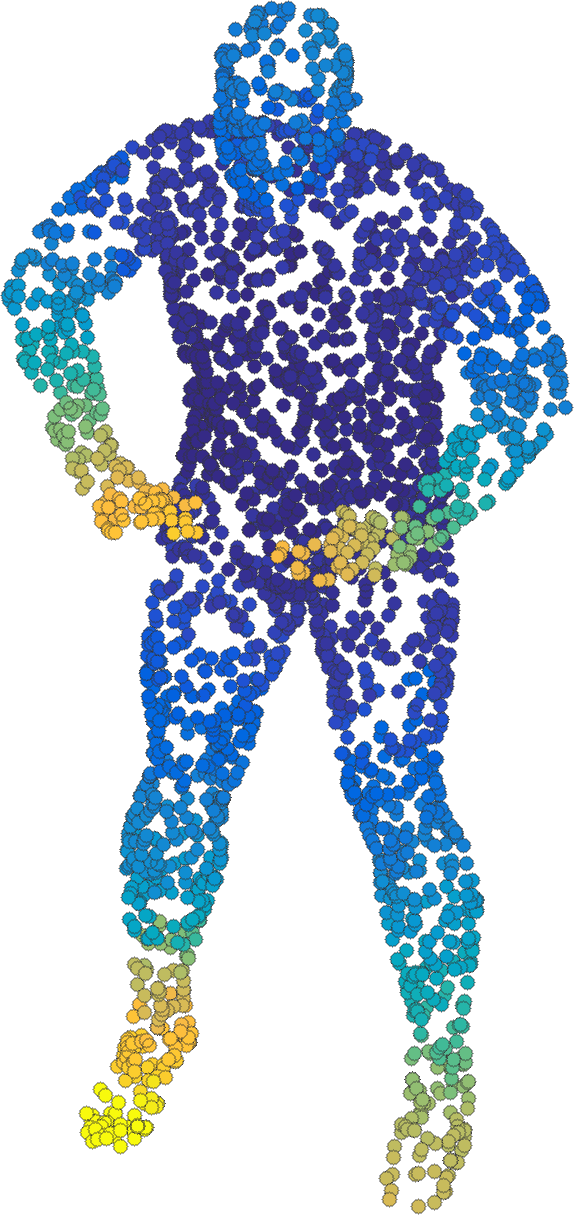} &
		\includegraphics[height=3.5cm]{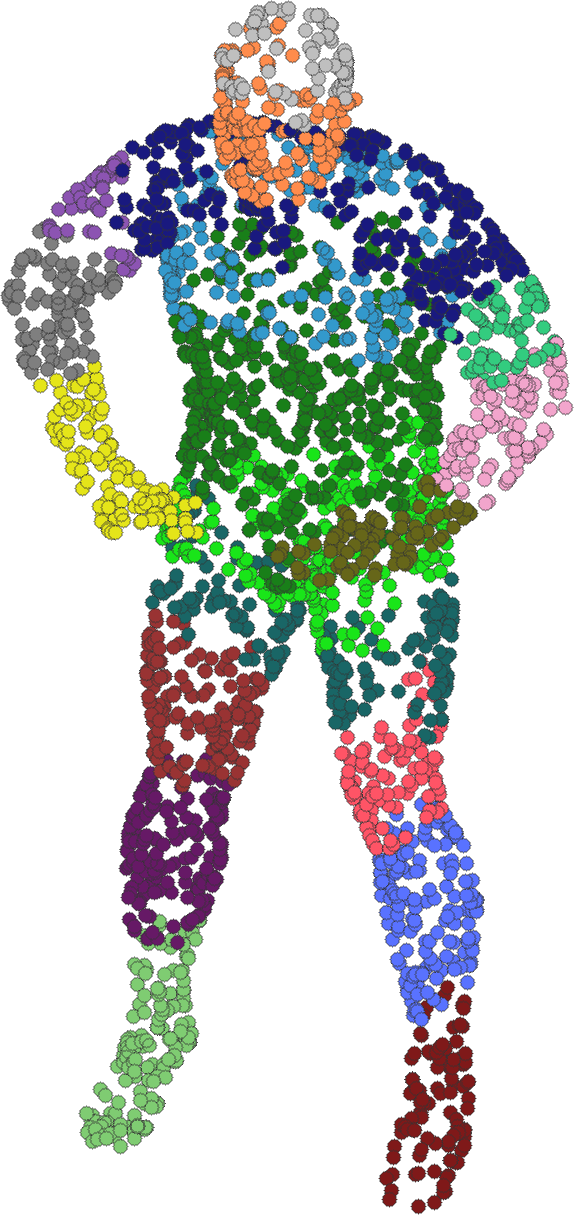} &
		\includegraphics[height=3.5cm]{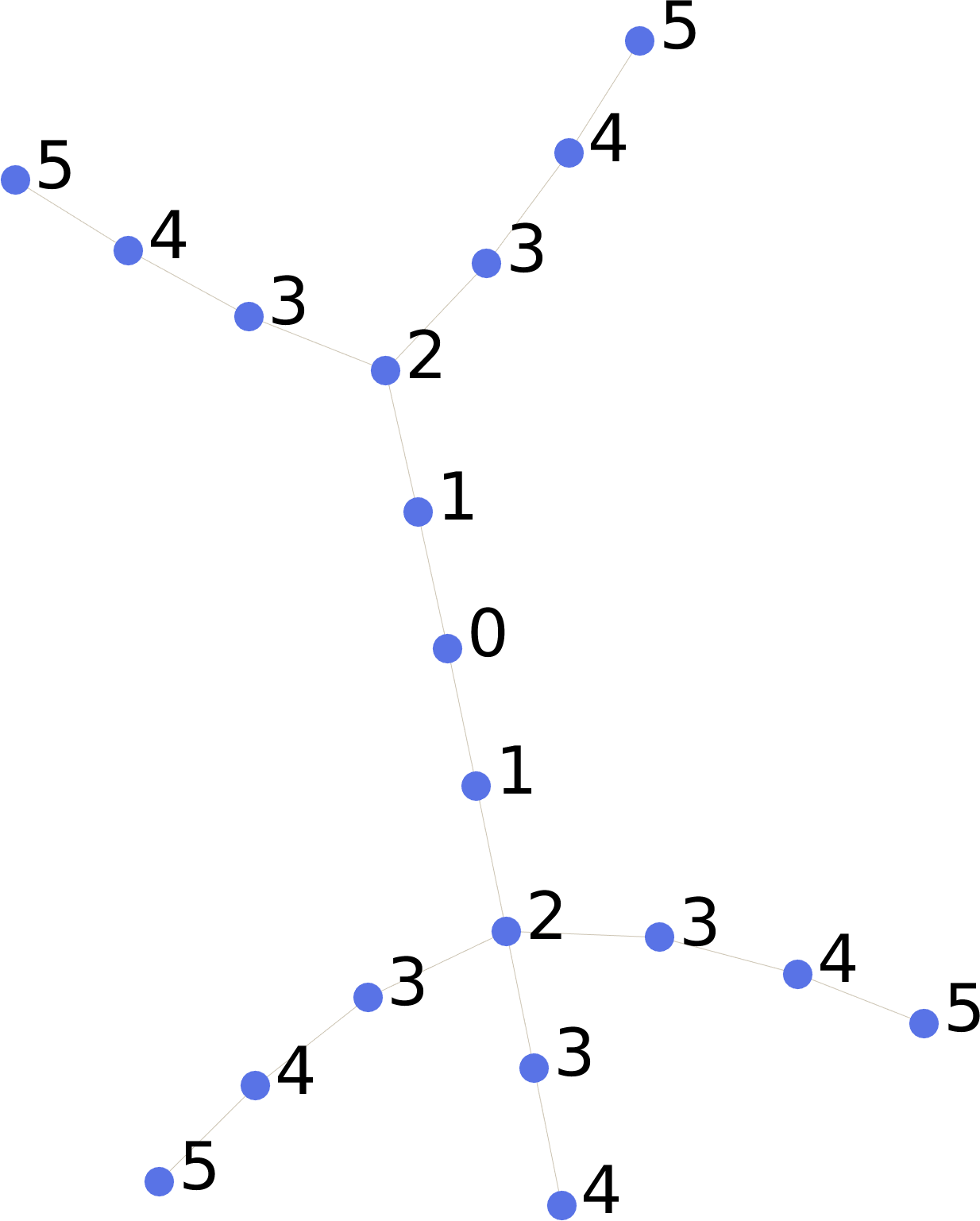} &
		\includegraphics[height=3.5cm]{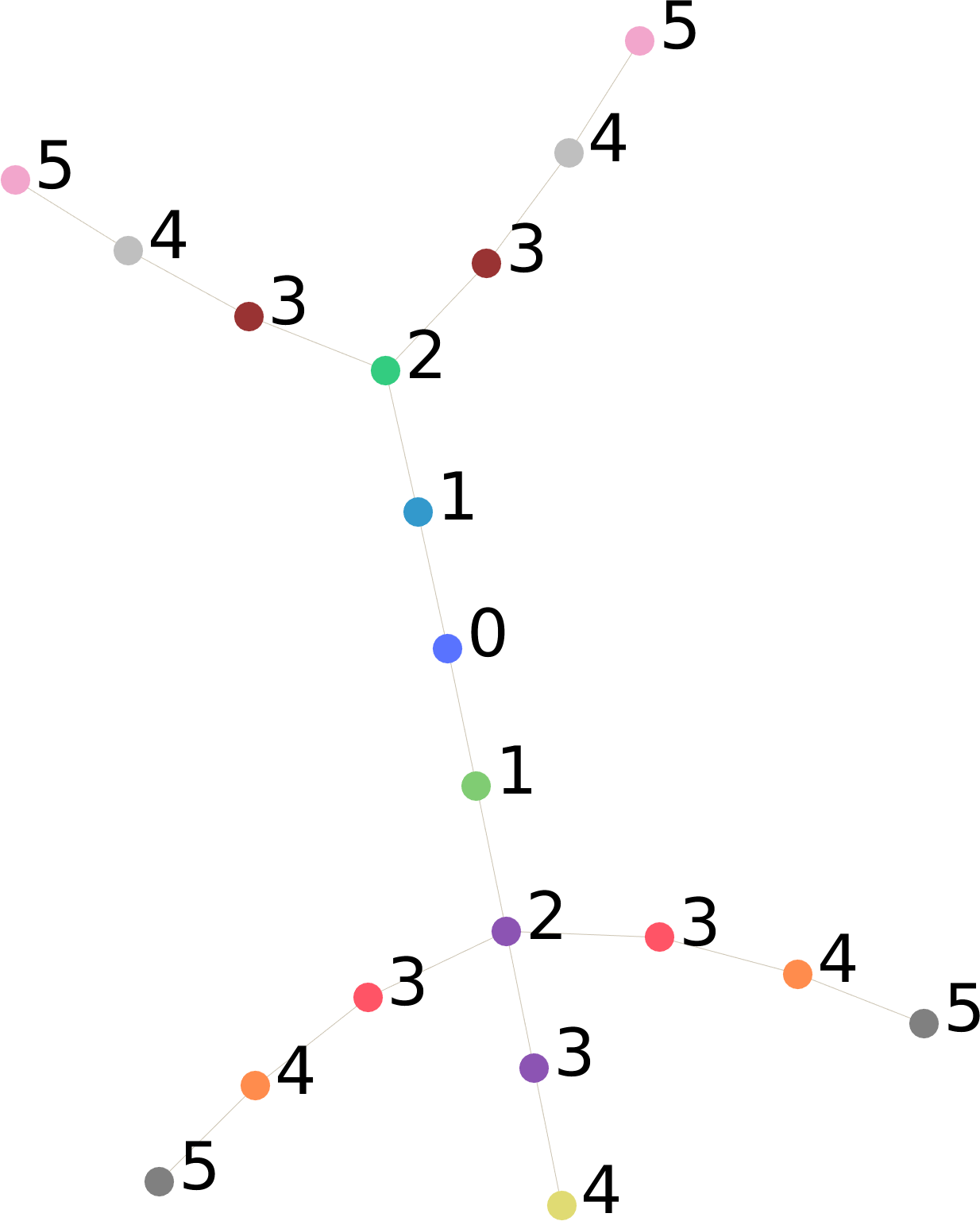} &
		\includegraphics[height=3.5cm]{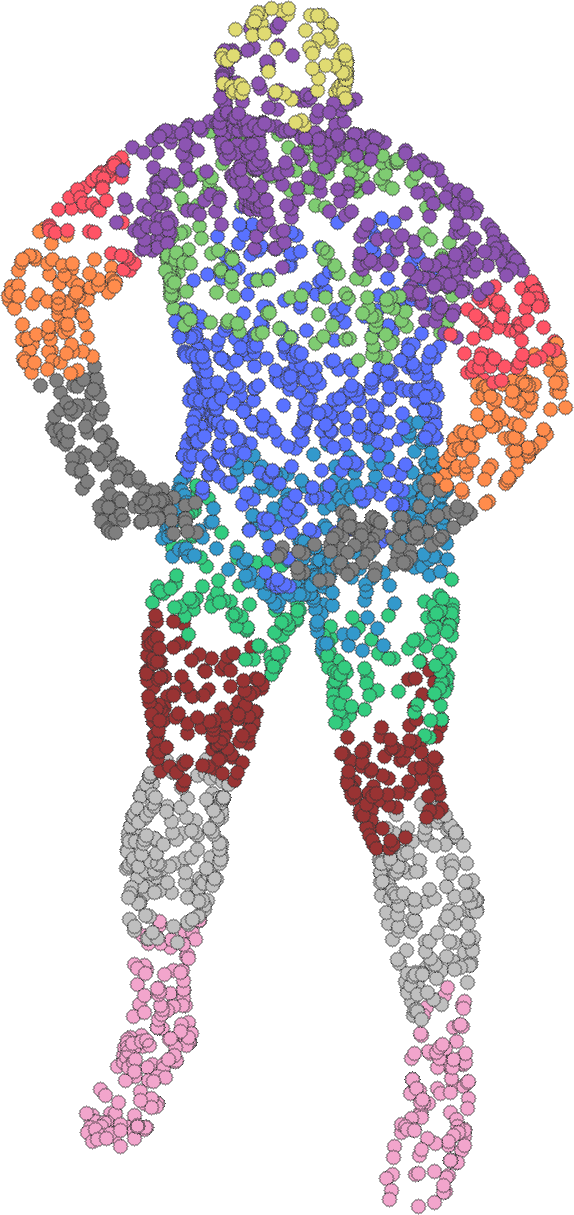} \\

		(a) & (b) & (c) & (d) & (e) \\

  \end{tabular}
\caption{Overview of our pipeline. We compute the HKS descriptor for the two shapes (a), then mutually segment them into consistent regions (b). A shape graph is created for each segmented shape (c). Then, the shape graphs are matched in a symmetric matching. In (d), matching nodes are shown with the same color. The matching between graphs induces a symmetric matching of the segmented shapes (e). \label{fig:symmetry_overview}\vspace{-5mm}}
\end{figure*}



\section{Related Work}



Methods for computing correspondences between shapes can be divided both according to the types of
input that they handle, and the kinds of output that they produce. Below we primarily consider
techniques for matching shapes undergoing \emph{non-rigid} correspondences, and especially those
that can handle significant variability and diverse shape representations. We refer the interested
reader to several recent surveys ~\cite{vankaick11,tam2013registration,biasotti15} and tutorials
\cite{eg11tutorial} for an in-depth review of this area.

Perhaps the most common representations for shapes in Computer Graphics and Geometry Processing are based on point
clouds and triangle meshes, and many methods for finding correspondences across such shapes have been proposed. Some
techniques are based on the assumption that shapes are approximately aligned and use local search (especially non-rigid
ICP) to find a better matching, e.g.  \cite{brown2007global,li2008global,chang2008automatic} among many
others. \change{Unlike these methods, our approach is global and does not require any spatial pre-alignment.}
\change{As such, our method is related to previous techniques for finding pose-invariant shape correspondences. Perhaps
  the most common general approach for addressing this problem} is based on preserving some prescribed geometric model,
such as approximate conformality \cite{lipman2009mobius,kim11}, or approximate intrinsic isometry (preservation of
geodesic distances), including \cite{bronstein2006generalized,coarsetofine,ovsjanikov10}, among many others.  Such
methods can work well when the deformation follows a prescribed model, but are often not able to handle significant
distortion or changes in representation.

More recently, a number of techniques have argued for establishing soft \cite{solomon2012soft} or
approximate correspondences between probability density functions, in part as a way to alleviate the
limitations of methods that seek point-to-point maps. \change{This includes both a large class of
  recent techniques based on the functional map representation (e.g.,
  \cite{ovsjanikov12,pokrass13,rodola2016partial,nogneng17} among others) and methods based on the
  formalism of Gromov-Wassterstein distances and optimal transport
  \cite{memoli2011gromov,solomon2016entropic}.}  \change{Such approaches can handle uncertainty and
  incomplete information more gracefully, but can still be computationally expensive, and often
  require additional information (such as the descriptor preservation constraints for functional
  maps \cite{nogneng17}) in order to obtain accurate results.} As we show below, our method can be
used in conjunction with these approaches to incorporate structural information into shape matching
to improve the final results.

Our approach is designed to obtain correspondences between regions on the
shapes. 
Several works have explored the related problem of shape co-segmentation~\cite{sidi11,kalogerakis12}.
Note that the goal of these techniques is typically to obtain a semantic high-level segmentation of
the shapes (i.e., to jointly decompose shapes into meaningful segments), whereas we aim to find
correspondences between smaller regions which might not correspond to semantically meaningful parts,
but, as we show, can enable multiple applications, including more accurate pointwise
correspondences.  \minorrev{We also note that our region-level correspondence method is related to
  robust shape segmentation techniques, e.g., \cite{rodola2014robust}, but unlike them takes a pair
  of shapes into account to produce explicit correspondences between regions. As such, it is similar to an
  approach by Pokras et al. \cite{pokrass2013partial} who optimize over matching parts
  jointly on a pair of shapes; however, we avoid the expensive indicator field optimization by using a simplified shape graph.}

Our method is perhaps most closely related to a recent technique of Ganapathi-Subramanian et al. \shortcite{ganapathi16}, who also concentrate on obtaining region-level correspondences between
shapes undergoing significant non-rigid deformations. Similarly to ours, their method is based on
the analysis of feature functions defined on a pair of shapes. However, we show how \emph{global
  shape structure}, in the form of a shape graph, can be extracted and used explicitly to obtain
accurate region-level matches. We compare our results extensively with \cite{ganapathi16} and show
that our method produces more accurate region correspondences, and is more efficient 
in both time and memory used.

We also note that our method of comparing shapes via their simplified graphs is related to skeleton-based shape matching
and retrieval \cite{sundar2003skeleton}. \minorrev{However, the matching
between skeletons does not directly induce a matching between points or regions on the shape. Producing a region to region matching from the matching skeletons can thus be a costly and error-prone process, with limited granularity.

Our method is also related to a recent technique that uses region-level comparison for quantifying shape similarity \cite{kleiman15shed}. 
However, in this work the regions are only matched implicitly as means to quantify similarity between shapes. Thus, this method is not suitable for matching shapes with large geometric distortions.}
\change{Moreover, we pay special attention to
the robustness of each part of our pipeline and use an intrinsic shape graph approach, with direct control of the graph
complexity based on an extension of the Mapper algorithm \cite{singh07}. Finally, our use of a robust set of shape
descriptors, and a joint shape decomposition, which yields strongly related regions, all together result in a method that is both
efficient and accurate even when comparing across different representations, such as triangle meshes
and point clouds.}

\change{Our symmetry-aware region-wise matching is related to methods that incorporate symmetry into
  the correspondence-finding pipeline, such as \cite{liu2012finding,tevs2014relating,ovsjanikov13},
  among others.} Our method is different in its efficiency, robustness and ability to handle
different shape representations, while also being useful downstream in pointwise matching
techniques.

\paragraph*{Contributions}
To summarize, our main contribution is a method for finding region-level matches between
non-rigid shapes that is:
\begin{itemize}
\item Robust in the presence of significant geometric variability.
\item Significantly more efficient than the most directly related approaches.
\item Capable of handling different representations, such as triangle meshes and point clouds.
\item Able to both produce matches that mix symmetric parts, or disambiguate between them, when necessary.
\end{itemize}


We evaluate our method extensively on a wide variety of existing benchmarks, in both a qualitative
and quantitative manner and show that the region-based correspondences produced by our approach are
both more accurate than those of most directly related methods, and can be used directly within
the existing point-to-point matching methods to obtain more precise correspondences, while being
both faster and more robust.




\section{Method Overview}


Our method consists of two main steps: generating a shape graph for each shape using a consistent joint segmentation of the two shapes, and matching the nodes of the shape graphs. In addition, we demonstrate an optional application of producing accurate point-to-point maps from the region-based correspondences.

Given a pair of shapes, we generate their shape graphs using an extension of the \texttt{Mapper} algorithm~\cite{singh07}.
First, we compute a shape descriptor on each shape, in the form of a set of real-valued functions. 
Ideally, the shape descriptor should capture the structure of the shape while being invariant to specific geometric details, and robust to noise. In our implementation, we use the HKS~\cite{sun09} functions for several time steps, which intuitively indicate how close each point is to an extremity or a high curvature area in the shape. \change{We jointly cluster the descriptor values of the two shapes, and construct a shape graph from the clusters. The shape graphs preserve only the structure of the shapes with no geometric information.} We describe the joint segmentation and construction of the shape graphs in detail in Section~\ref{sec:segmentation}. The matching between shape graphs is computed using an adapted spectral correspondence technique, which is described in Section~\ref{sec:matching}.

Figure~\ref{fig:symmetry_overview} provides an overview of our pipeline.  In (a), the shapes are colored according to the value of a single function. Next, we use the shape descriptors to jointly segment each shape into regions (b), and generate shape graphs where each node corresponds to a region on the shape (c). Then, a correspondence between the shape graphs is computed (d), inducing a correspondence between the regions of the two shapes (e). The symmetric correspondence can then be used to find a one-to-one correspondence between regions when necessary, and used in a variety of settings including computing accurate point-to-point maps.


To compute point-to-point maps between the shapes, we use the functional maps framework of~\cite{ovsjanikov12}, which can easily accommodate segment correspondence constraints to obtain point-to-point maps. The segments used in~\cite{ovsjanikov12} are sparse, whereas we provide a dense set of segments which cover the entire shape, which leads to significant improvement in map quality.

\section{Consistent Segmentation}
\label{sec:segmentation}

The input to our method is two shapes, which can be represented as triangular meshes or point clouds. 
The first step is to co-segment these shapes in a consistent manner, and to build their corresponding shape graphs. 
By consistent we mean that the edges between segments have a similar structure in both shapes, with respect to some underlying (unknown) map.
The consistency of the segmentation is important for two reasons. First, a consistent segmentation implies that similar
shapes produce similar shape graphs which can be better matched; in particular, isomorphic shape graphs can be matched
in a bijective way. Second, a matching between consistent segments induces a more accurate matching between vertices.
Note that we do not seek a semantic segmentation of the shape; a semantic part may be cut into several segments.

\change{Many works represent shapes as graphs of segments. Reeb graphs are sometimes used as a skeletonization of the
shape~\cite{gebal09,barra13}. However, Reeb graphs are sensitive to small features in the shape and are therefore not
ideal for representing the structure of a shape in a robust and consistent way (see also Fig. \ref{fig:reeb} below).} Our approach is similar to
\texttt{Mapper} graphs \cite{singh07}, a related but more stable construction which is essentially a discretized
generalization of the Reeb graphs. 

\change{

More specifically, suppose we are given a pair of shapes $M$ and $N$ and a $d-$dimensional feature descriptor on each. Such feature descriptors can be thought of as either a set of $d$
real-valued functions $f_{i}^M$ and $g_{i}^N$, $i=1..d$ or as two point-clouds in $d$-dimensional
space, where each point corresponds to the values $\{f_i(x)\},$ $i=1..d$ for some fixed $x$ on
$M$ or $N$. Our graph construction procedure then proceeds as follows:
\begin{enumerate}
\item For each dimension $i$, we align the range of the values of $f_{i}^M$ and $g_{i}^N.$ This step
allows us to match shapes with large isometric distortion as shown in
Section~\ref{sec:symmetry_evaluation}. For this, we first generate a set of mappings $T_i: \mathbb{R}
\rightarrow \mathbb{R}$ between the function values of $f_{i}^M$ and $g_{i}^N.$ We order the vertices of the
two shapes according to their function values, and map the function values of the second shape to
the values with the matching \emph{rank} in the first shape. To support shapes or point clouds
with a different or non-uniform sampling, we define the rank of a function value as the area
of the points on the shape that have a lower value. More precisely:
$$T_i(\beta) = \{\alpha : \text{Area}_{M}(x : f^M_i(x) \leq \alpha) = \text{Area}_{N}(y : g^N_i(y) \leq \beta)\}.$$
We then use $T_i$ to replace
$g_i^N$ with $h_i^N = T_i \circ g_i^N$.
\vspace{1mm}
\item We construct the point cloud $P$ in $d$-dimensional space obtained by merging point clouds
  obtained from $\{f^M_i\}$ and $\{h^N_i\}$. The number of points in $P$ equals to the sum of the
  number of points in $M$ and $N$. We then perform $k$-means clustering on $P$ to obtain a set of
  $k$ centroids $c_j \in \mathbb{R}^d, j = 1..k$, and the associated shape decomposition into sets
  $C^M_j = \{x \in M : c_j \text{ is nearest to } \{f^M_i(x)\} \}$. Each set $C^M_j$ may contain several
  connected regions on $M$, for example regions on both arms and both legs which might have similar
  descriptor values.
\vspace{1mm}
\item We then construct our shape graph by first creating a node for \emph{every connected region}
  of every set $C^M_i$. Then, for each connected region $R$ associated
  with the centroid $c_i$, we create the ``expanded region'' $R^*$ by computing the
  connected region of
  points on $M$, which contains $R$, and for which $c_i$ is either first \emph{or} second nearest centroid. Finally, we
  create edges between the node corresponding to $R$ and every node whose region $R^*$ overlaps.
  Here, our use of second-nearest centroids can be thought of as an extension of the overlapping segments in \texttt{Mapper}
  to high-dimensional range functions.
\vspace{1mm}
\item We construct the shape graph associated with shape $N$ in the same way by decomposing it into sets $C^N_i$
  and performing the previous step, with the points on $N$.
\vspace{1mm}
\item Finally, we run steps 2-4 above for a range of $k$ (between 5 and 10) in $k$-means 
and pick the result so that the two shape graphs are as similar as possible, where we measure similarity by comparing the histogram of node degrees of the graphs. 
\end{enumerate}
The output of this procedure is a pair of graphs, one for each shape $M$ and $N$ such that each point on the
shapes is associated with exactly one node on the corresponding graph, and moreover the graph nodes
are labeled by the index $j$ of the centroid in feature space. These indices are consistent (i.e.,
can be compared) across nodes on the two graphs, although we do not use this property in our
matching pipeline, described in Section \ref{sec:matching}.

}

We remark that the construction above closely follows the construction of \texttt{Mapper} graphs
proposed in \cite{singh07}. However, rather than tiling the parameter space using bins or pre-defined
tiles as suggested in \cite{singh07} (Section 3.1, Example 3.3) we consider the tiling produced by
pairs of adjacent Voronoi cells, given by the centroids of $k$-means clustering. This allows us to
better reflect the parameter space and relations between different functions.
\change{Moreover, our final graphs can be thought of as dual to the Mapper graphs since each point
  on the shape is associated with exactly one node on the graph.}
Nevertheless, one of our motivations is to demonstrate the utility of \texttt{Mapper} graphs for
shape correspondence problems, inspired by recent results of stability \cite{carriere15} under
certain perturbation models. In our setting, these graphs are especially useful for two reasons:
first, by controlling the number of clusters, we gain precise control on the complexity of the
resulting graph, i.e., the number of nodes it will have. Second, by using large overlapping regions
we obtain significant robustness with respect to perturbation of function values. Finally,
constructing a \texttt{Mapper} graph can be done extremely efficiently (in linear time) as only
computation of connected components is required, which contributes to the overall efficiency of our
pipeline.


For the descriptor function, we use the \emph{heat kernel signature} or HKS~\cite{sun09,gebal09} for a given range of time steps. 
In our experiments we used $15$ time steps between $t=0.03$ and $t=0.25$ (the shapes are normalized to have unit areas).
We find that the HKS descriptor is very robust to changes in geometry or articulation,
with the lowest HKS value located in the central region of the shape and the highest HKS values at the shape's
extremities. However, other descriptors can be used instead or in conjunction with HKS.
\minorrev{These descriptors should be smooth, robust, and descriptive. Smoothness is essential so the segments are not overly fragmented. The descriptor should be robust enough to be able to match shapes with varying geometry, while still being descriptive enough to differentiate between different elements of the shape. For example, we experimented with the Wave Kernel Signature~\cite{aubry11} and found that while it can be more informative, the HKS is more robust under shape change and results in less noisy shape graphs.}

In the case of triangle meshes, we use the classical cotangent-weight discretization to compute the HKS function. For
point clouds, we follow the general approach of \cite{Belkin09} and define a Laplacian with a Gaussian weight on a
$k$-nearest neighbor graph, where the variance is fitted by considering the mean edge length. We note that although
the individual eigenfunctions of the Laplacian might be unstable across representations, the HKS for a large time value
is remarkably resilient with respect to changes in sampling density and representation. Moreover, as we ultimately use
the induced Mapper graph for matching rather than the values of the function itself, we gain an additional level of
robustness.


\change{
\paragraph*{Perturbation.}
The motivation behind step 5 in our pipeline described above is that
small geometric details in the shapes may sometimes cause descriptor values to differ such that
extra segments in shape graphs are generated for some selections
of number of clusters (and not for others). This is also suggested by the analysis of \cite{carriere15} as some shape
features may have borderline function values. To reduce this effect, we vary the number
of clusters between $5$ and $10$ (as fewer than $5$ clusters yield regions which are too large and
uninformative, while more than $10$ clusters yield segmentations which are usually too noisy). For
each setting we compute the shape graphs for both shapes, and evaluate the similarity between the graphs by comparing the histograms of node degrees in the two graphs. We consider
the graphs with the least differences between the lists as the most similar (isomorphic graphs will
produce identical lists). Finally, we select the number of clusters that produce the best match
between shape graphs.}


The combination of using a stable descriptor (HKS), joint segmentation, and searching for
segmentations with similar structure generates consistent segmentations for shapes from the same set
(for example two human shapes) as well as shapes from different sets, such as matching a cat to a
dog or a man to a gorilla. An example of consistent segmentation is given in
Figure~\ref{fig:symmetry_overview}(b), with the corresponding shape graphs shown in
Figure~\ref{fig:symmetry_overview}(c).

\section{Region Matching}
\label{sec:matching}


Once similar shape graphs are computed, our goal is to find a matching or correspondence between
them. This is made challenging for two reasons: first, the shape graphs that we obtain are not
necessarily 
isomorphic, and thus our approach must be able to deal with partial and approximate information.
Second, many shapes in practice (both man-made and natural), and the majority of shapes that we
consider exhibit different symmetries, which means that often multiple equally good solutions might
exist to the graph matching problem. The most common symmetry in our case is reflectional left-right
symmetry, and thus we adopt a method capable of dealing with these challenges.
 
For this, we propose a two-step solution that first finds a symmetric correspondence between regions
and then breaks the symmetries. The symmetric correspondence factors out symmetries by allowing
groups of regions in the first shape to be matched to groups of symmetric regions in the second
shape. Thus, the direct and inverted maps are merged onto the same map. This relaxation of the
one-to-one correspondence constraint enables the use of an extremely quick and efficient algorithm.
In the second step, we break the symmetries using a simple algorithm to provide one-to-one
correspondence between shape regions.

To compute the symmetric matching, we rely only on the graph structure of the shape, without
additional geometric data. That is, the only information that we use when matching graphs are their
connectivity and node degrees. Therefore, symmetric branches cannot be distinguished from each other.
Branches which are not symmetric can be distinguished by their position in the shape graph. The nodes
of two symmetric branches (for example two legs) have the exact same properties, while non-symmetric branches 
(e.g. an arm and a leg) can be distinguished by the position of the head, even if each branch has the same number of
nodes.

\vspace{-1mm}
\subsection{Subspace-based Spectral Matching}
There exist several techniques for finding approximate solutions to graph correspondence problems
for shape matching, including various approaches based on quadratic programming relaxations, e.g. as
proposed by Berg et al.~\cite{berg05}, or more recently by Kezurer et al. \cite{kezurer15}. As we
target a time and memory-efficient approach, we use the spectral matching method of Leordeanu and
Hebert~\cite{leordeanu05}, which we adapt to our setting.

Namely, given two graphs with $n$ and $m$ number of nodes respectively, we construct a matrix $M$ of
$n \cdot m$ rows and $n \cdot m$ columns. This matrix contains the values of the unary
(first-order) terms on the diagonal, which can be interpreted as the affinity between a pair of
nodes in the two graphs. In the off-diagonal entries we store the second-order affinities between
\emph{pairs} of matches of nodes. Thus, the second-order terms describe the compatibility between a
match between two nodes $(i, j)$ and another match between nodes $(k, l)$. 
Finding a binary one-to-one correspondence vector $x$ that maximizes $x^T M x$ amounts to
solving the quadratic assignment problem, which is known to be NP-hard. Instead, the spectral
matching technique \cite{leordeanu05} relaxes the binary constraint and instead uses the eigenvector
corresponding to the largest eigenvalue of $M$, which is then discretized to find a solution.

In our solution, we define the unary term and binary term as follows.
For the unary cost of node $i$, we compute the histogram of graph distances, $H(i)$, by counting how many nodes have a graph distance of $1$ (adjacent), $2$, $3$, and so on to node $i$. This forms a vector which signifies the connectivity of the node.
Our first-order cost for a match between nodes $i$ and $j$ on the two graphs is then given as follows:
\begin{equation}
\label{eq:symmetry_unary_cost}
	C(i, j) = \norm{H(i) - H(j)}.
\end{equation}
The affinity between the segments is computed by:
\begin{equation}
\label{eq:symmetry_unary_affinity}
	U(i, j) = exp(-C(i,j) / \sigma),
\end{equation}
where in our experiments $\sigma = 0.5$.

For the second-order term, we compute the difference of graph distances:

\vspace{-7mm}
\begin{equation} 
	d_g(i, j, k, l) = \norm{g(i, j) - g(k, l)},
\end{equation}
where $g(i, j)$ is the graph distance between nodes $i$ and $j$. We also use
the difference of the unary costs between the matches:
\begin{equation}
	d_u(i, j, k, l) = \norm{C(i, j) - C(k, l)},
\end{equation}
where $C(i, j)$ is as defined above.
The second term is helpful for matching partial shapes or shapes with missing parts since the missing parts affect nearby segments in a similar way, which is reflected by this term.
Again, the affinity between the two matches is computed by:
\begin{equation}
\label{eq:symmetry_binary_affinity}
	U(i,j,k,l) = exp(-(d_g(i,j,k,l) + d_u(i,j,k,l)) / \sigma),
\end{equation}
with $\sigma = 0.5$.


\minorrev{In addition to the terms above, we have experimented with terms
that take into consideration the joint cluster id associated with the regions, produced during joint segmentation process, as described in Section \ref{sec:segmentation}. In our experiments, we found the results to be very similar in the vast majority of cases. Nevertheless, we recognize this might be useful in certain challenging cases and include this as an option in the public release of the code. The results shown below were computed without such terms.}

Our key observation in this section is that under some conditions, the continuous optimizer $x$,
corresponding to the largest eigenvalue of $M$ is a sparse vector. Namely, as the second-order costs
are diminished, $x$ becomes more and more sparse. In the general case, $x$ will converge to match
the highest value in the first order affinities. In our case, the first order costs contain many
repeated values, which correspond to possible matches, in particular because the nodes in our graphs
are associated with \emph{integer} values, arising from the discrete nature of the graph properties.
However, these first-order matches might include matches that are incorrect, such as
matching an arm to a leg or a head to a tail.

Our goal is to restrict the eigenvector to a specific subspace which spans only the correct solution
(up to intrinsic symmetry), while keeping it sparse. 
To this end, we use the second-order data as a \emph{tie-breaker}.
It directs the optimization towards a specific solution within the
subspace of possible solutions spanned by the first-order data, 
while staying within that subspace.
We thus scale the second-order data to a small value, for example by dividing it by the number of
non-zero elements in the matrix $M$. This results in a sparse vector within the subspace, which
respects the second-order data as well as the first-order data, and can be easily discretized.

To summarize, our algorithm consists of the following steps:
\begin{itemize}
\itemsep0.2em 
\item Compute the first-order and second-order data terms.
\item Scale the second-order terms to a small fraction, for example by dividing them by the number of non-zero elements in the matrix.
\item Compute the first eigenvector $x$ of the affinity matrix $M$.
\item Discretize the sparse eigenvector.
\end{itemize}
\vspace{5pt}

\change{ The entries in the eigenvector can be interpreted as the
  likelihood of each match to be part of the optimal solution. The
  construction of the first-order and second-order terms above ensures
  that symmetric regions have nearly the same likelihood.  Therefore,
  for each region we search for the most likely matches, while
  allowing a few symmetric matches to be selected up to a predefined
  maximum symmetry order, provided that their likelihood is similar.
  To this end, we search for a significant gap in the likelihood
  values.  If the gap occurs before the maximum symmetry order, we
  consider the matches before the gap to be correct. A gap that occurs
  after the maximum symmetry order indicates that the region does not
  have a strong match in the target shape, and the region is left
  unmatched.  In all of our examples, we allow the maximum symmetry
  order to be $8$.  To further eliminate inconsistent matches, we
  perform the same process by reversing the roles of the source and
  target shapes and only output matches which are selected for both.

To summarize, the discretization process is as follows:
\begin{itemize}
\item For each node in the graph, sort the values in the corresponding 
portion of the eigenvector from high to low.
\item Search for the first gap in the values: 
a value which is less than $90\%$ of the previous value.
\item If this gap occurs on value $i$ before the maximum symmetry order, 
then pick all matches that correspond to the first $i-1$ values.
\item Otherwise, consider this a low confidence match, and leave the part unmatched.
\item Perform the same process for both shapes and only output matches which appear
in both directions.
\end{itemize}
}

An example of the matching between graphs is shown in Figure~\ref{fig:symmetry_overview}(d). Nodes
with the same color are matched to each other. The induced matching between shape segments is shown in Figure~\ref{fig:symmetry_overview}(e).

\subsection{Symmetry Breaking}
\label{seg:sym_breaking}

\change{
The goal of our symmetry breaking process is to produce a one-to-one matching
between the shapes which is consistent and without discontinuities.
Similarly to previous methods, the produced one-to-one matching may be
consistently flipped, for example by mapping the entire left side of a human 
to the right side. In our evaluation, we do not penalize maps that are 
exact symmetric flips with respect to the correct one, similarly to previous
work (see~\cite{kim11,ovsjanikov12}).

To produce a one-to-one matching, we use a simple heuristic which is relatively
quick to compute, yet produces the correct result in the majority of cases. 
This demonstrates that decoupling the symmetry breaking from the shape 
matching process effectively simplifies both problems.
The symmetry breaking is based on the assumption that geodesic distances 
between nearby regions do not change drastically between the shapes.
Therefore, at each step we match regions which are nearest to 
previously resolved regions.

Our symmetric shape matching produces sets of matching regions, where $k_1$ 
symmetric regions in the first shape match $k_2$ symmetric regions 
in the second shape.
We start by selecting a single set of symmetric matches where $k_1>1$ and
$k_2>1$ are as small as possible. For simplicity, assume the typical case
of $k_1, k_2 = 2$, where regions $R_1, R_2$ from the first shape match regions
$S_1,S_2$ from the second shape.
We produce a one-to-one matching for the group arbitrarily, for example by
matching $R_1$ to $S_1$ and $R_2$ to $S_2$. 
We define groups $V$ and $W$ of vertices that belong to $R_1$ and $S_1$ respectively.
In each subsequent iteration, we resolve a region $R_i$ of the first shape 
which is matched to more than one region and has the minimal distance to 
a vertex in $V$, until all regions are resolved.
To this end, we compute the average geodesic distance from each region in the 
first shape to $V$ and 
pick the region $R_i$ minimizing this distance.
We then proceed to find the region $S_i$ which is part of the group of matches
of $R_i$ and has the minimal distance to a vertex in $W$.
We set $R_i$ to match $S_i$ and remove all other matches of $R_i$ and $S_i$.
Finally, we add the vertices of $R_i$ and $S_i$ to $V$ and $W$ respectively.}
This process is used to subsequently produce point-to-point maps and the results in Table~\ref{tbl:p2p} and Figure~\ref{fig:faust_fmaps}.

\section{Evaluation}\label{sec:symmetry_evaluation}

\begin{figure}[t]
	\centering
	
    \includegraphics[height=5cm]{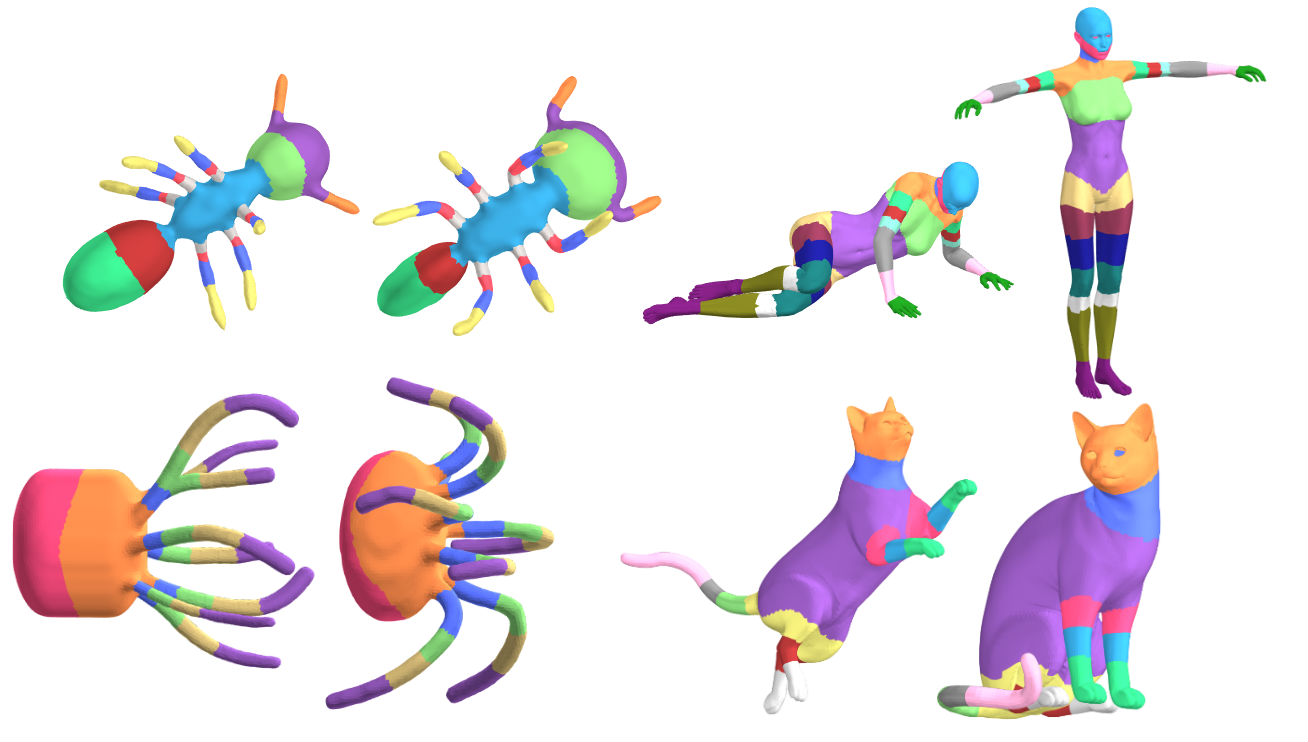}	
    \caption[Region-based correspondence between nearly isometric shapes]
    {\change{Region-based correspondences between nearly isometric shapes. For
    each pair, regions with the same color are matched.}
	\label{fig:gallery_iso}
	\vspace{-4mm}}
\end{figure}



In this section, we evaluate four main aspects of our method: the accuracy of the region matching
between shapes, the robustness of our method on point clouds, the point-to-point maps generated
using our region-based correspondence, and an application for finding symmetric regions in a single
shape.

In Section~\ref{sec:qualitative}, we first provide intuition to the types of transformations our
method can handle with a few qualitative results. We also show qualitative examples of finding symmetric regions by matching a shape to itself. In Section~\ref{sec:region}, we compare the
accuracy of our region-based matching to BIM~\cite{kim11} and the recently-proposed stable regions method of~\cite{ganapathi16}. 
\change{In addition, we compare our method with a baseline approach by
  replacing our shape graph construction with classical reeb graphs
  and matching them instead.}
In section~\ref{sec:pc_evaluation}, we analyze the robustness of our
approach for point clouds with varying sampling sizes and noise levels. We also evaluate the
matching between 
triangle meshes and point clouds and compare them
to~\cite{ganapathi16}. In Section~\ref{sec:p2p} we generate point-to-point maps using the functional
maps framework, and compare them with~\cite{kim11,ovsjanikov12,ganapathi16}. We show that our method
is more robust and informative than \cite{ganapathi16} and leads to more accurate point-to-point maps.
Finally, in Section~\ref{sec:timing}, we discuss the running time of our method compared 
to existing methods.

\change{
\paragraph*{Datasets.}
We use the following well-known shape datasets for evaluation. TOSCA~\cite{tosca} contains 
$80$ shapes in $9$ categories, where each category contains the same shape in different poses,
so that isometric distortion within a category is typically low. The models of each category are
registered with a mapping between vertices that we use to evaluate the correctness of our
correspondences. SCAPE~\cite{scape} contains $71$ shapes of the same person in different 
poses with low isometric distortion. These shapes also have a mapping between them.
MPI FAUST~\cite{faust} dataset contains $100$ shapes of $10$ different human models in
$10$ poses. The models vary in body type and they way they assume each pose so there is 
a lot of isometric distortion between models of different categories. However, all shapes
in FAUST have a mapping between them, which allows us to numerically evaluate correspondence
between shapes with high distortion.

\begin{figure}[t]
	\centering
	
    \includegraphics[height=5cm]{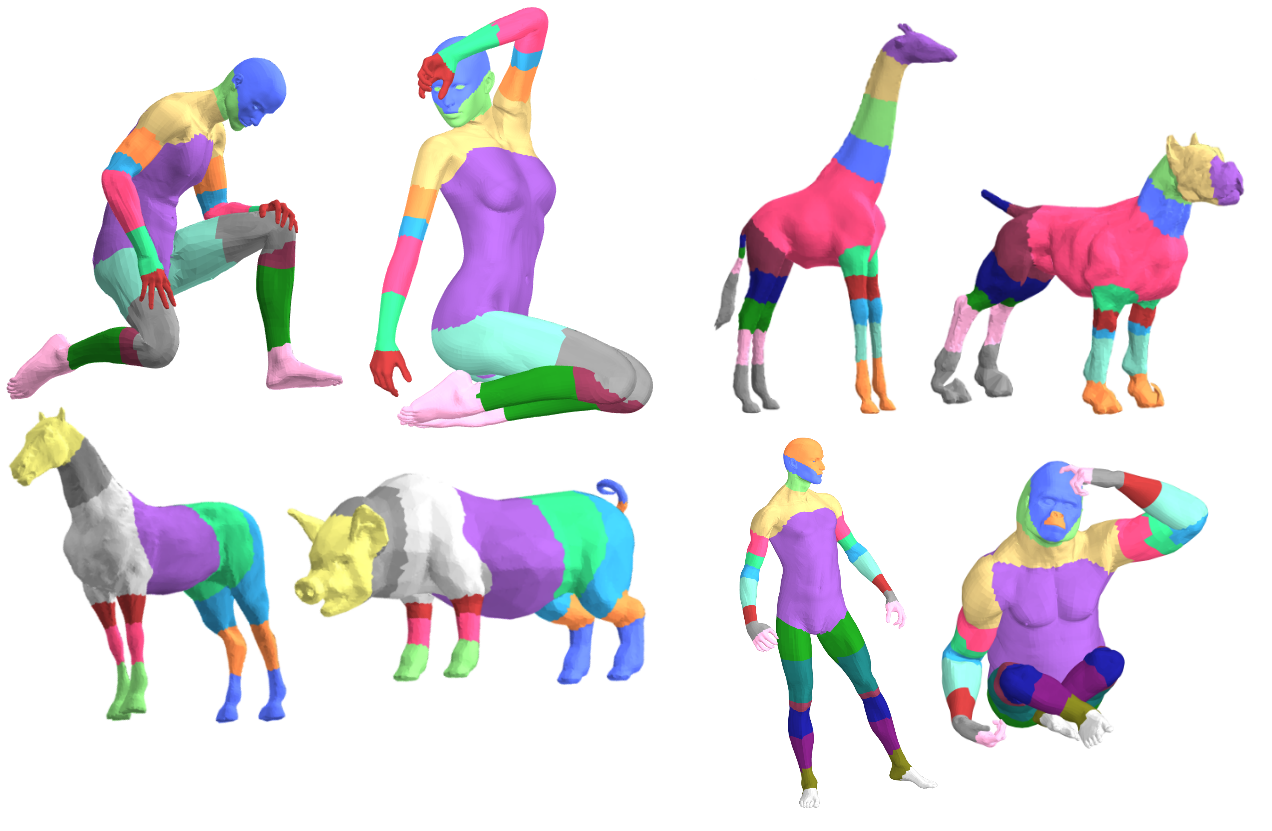}
	 \caption[Region-based correspondence between non-isometric shapes]
    {\change{Region-based correspondences between pairs of shapes with significant isometric distortion.}
	\label{fig:gallery_noniso}
	\vspace{-4mm}}
\end{figure}

For each of these datasets, we randomly generated a list of pairs such that each shape
is matched with one random shape from the same category. For FAUST, we augmented the set
of pairs to include shapes from different categories, with the same pose or different pose.
In total we evaluated $244$ pairs of shapes from FAUST dataset, which include $111$ pairs of
the same category (i.e. same model), $20$ pairs from different categories but the same pose,
and $113$ pairs from different categories and different poses.

For qualitative evaluation, we include a few examples of matching
shapes from the Princeton Segmentation Benchmark~\cite{chen09} (PSB),
which includes shapes in different classes, but for which we do not
have a reliable ground truth correspondence.  }



\begin{figure}[t]
	\centering
	
    \includegraphics[height=6.5cm]{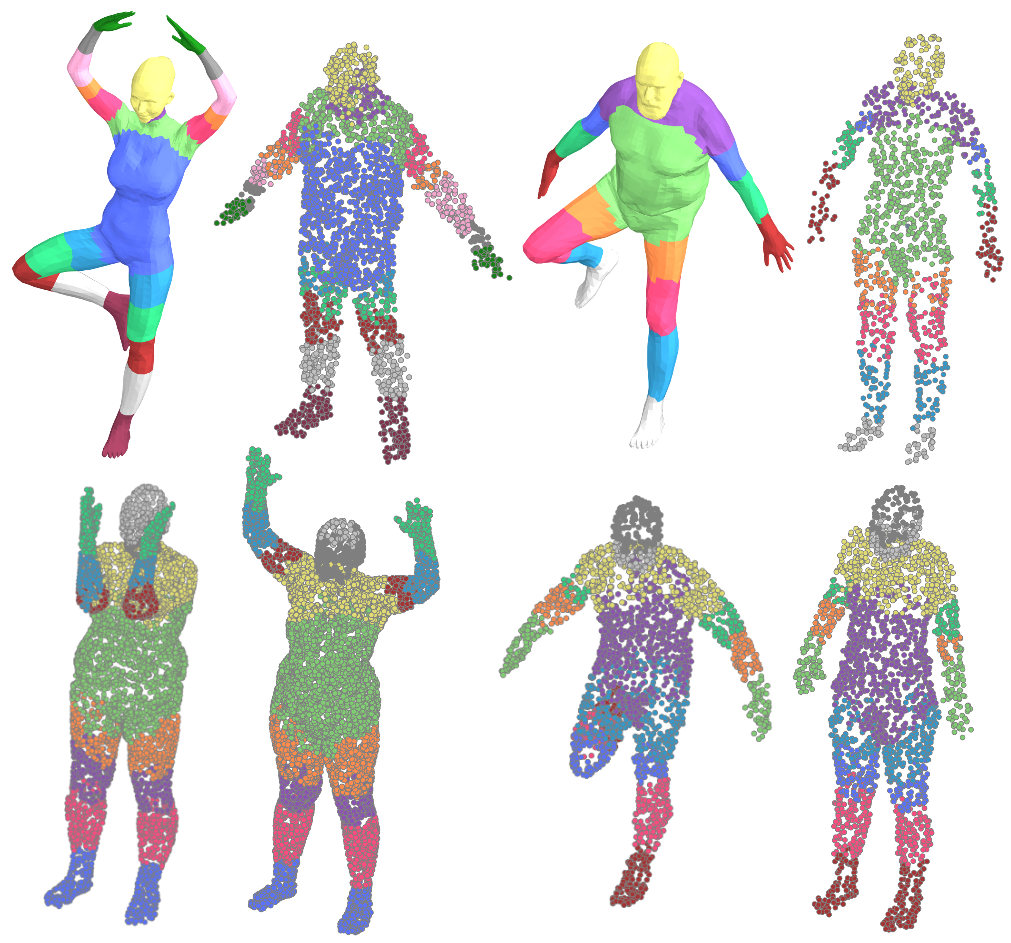}
	
    \caption[Region-based correspondence between point clouds]
    {\change{Region-based correspondence between triangular meshes and point clouds (top row), and between point clouds (bottom). 
    The shapes vary in pose, body type, sampling density, and sampling noise.}}
	\label{fig:gallery_pcd}
	\vspace{-3mm}
\end{figure}

\subsection{Qualitative Evaluation}
\label{sec:qualitative}
\begin{figure}[t]
	\centering
	
    \includegraphics[height=6.5cm]{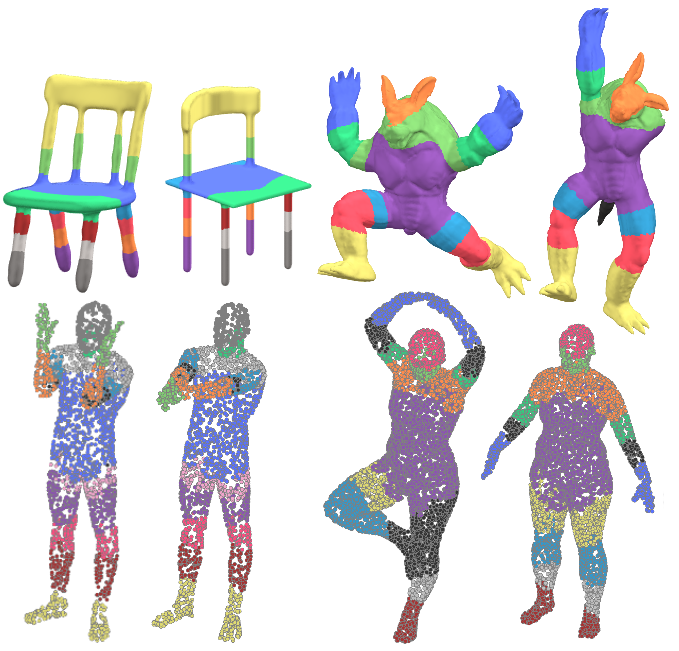}
	
    \caption[Region-based correspondence between shapes with topological noise]
    {\change{Region-based correspondences between shapes with different
      topological structure. Unmatched regions are in black.}}
	\label{fig:gallery_top}
	\vspace{-3mm}
\end{figure}

\begin{figure}[t]
\centering

	\begin{tabular}{@{}c@{\,}c@{\,}c@{}}

		\includegraphics[height=3cm]{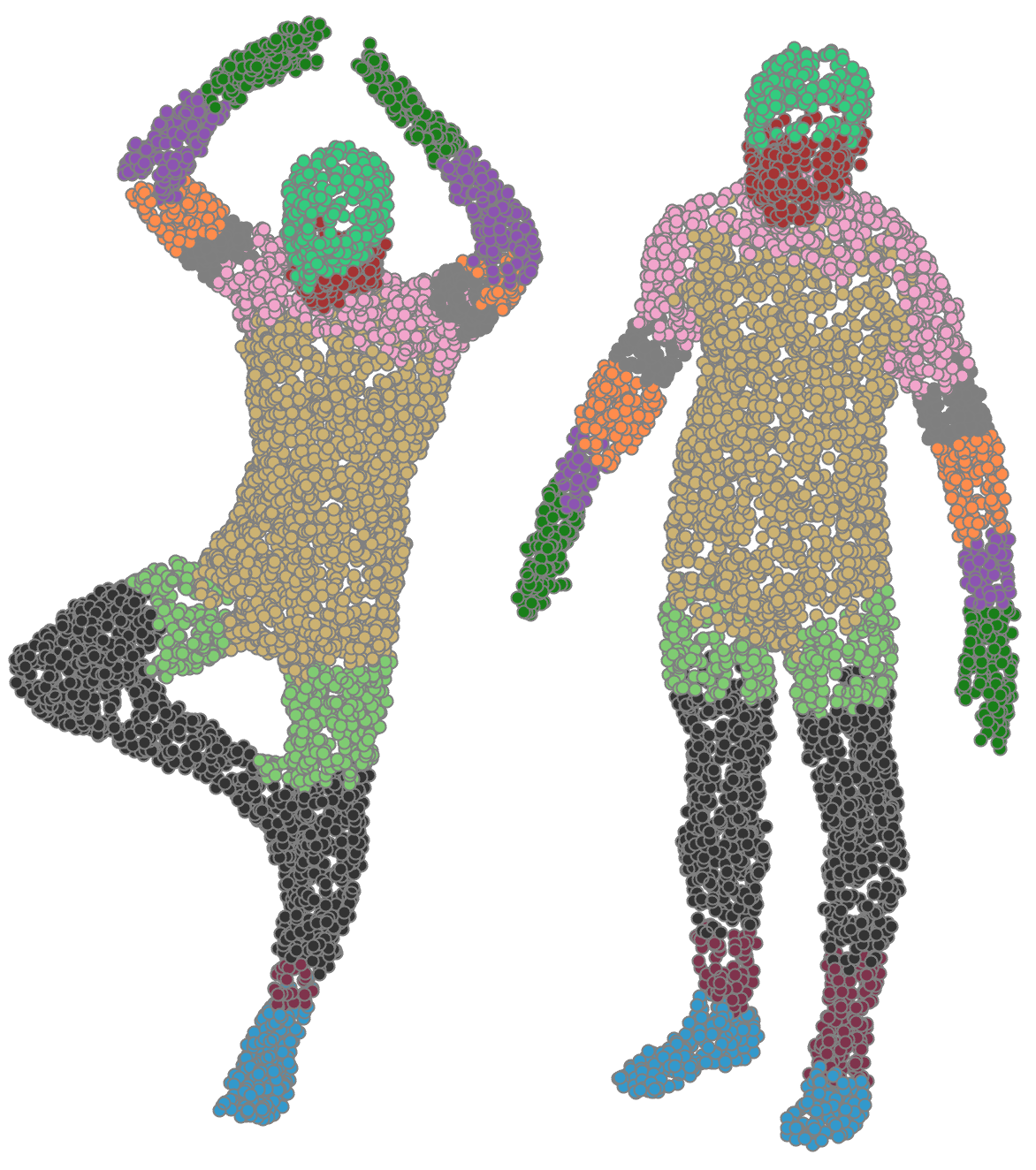} &
		\includegraphics[height=3cm]{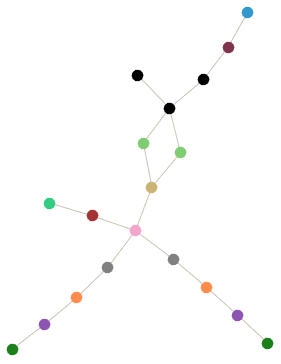} &
		\includegraphics[height=3cm]{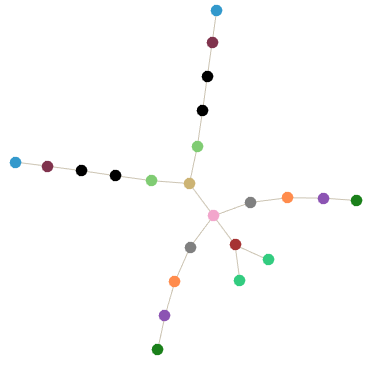} \\

	\end{tabular}	

        \caption[Matching shapes with non-isomorphic shape
        graphs]{\change{Region-based correspondence between shapes with
          non-isomorphic shape graphs with different topology.  The
          shape graph of the left shape (shown in center) contains a
          loop while the shape graph of the right shape (shown right)
          has a tree structure. \vspace{-2mm}}}
\label{fig:symmetry_non_isometric}

\end{figure}

\change{ To start the discussion of our results, we show examples of
  the region-level matches obtained by our method on various shapes in
 Figures~\ref{fig:gallery_iso},~\ref{fig:gallery_noniso},~\ref{fig:gallery_pcd} and~\ref{fig:gallery_top}. 
  For each pair, similar colors denote matching segments.
  Figure~\ref{fig:gallery_iso} shows typical results for shapes from
  the same category which are nearly isometric. Our segmentation
  method produces small isomorphic shape graphs, which are easy
  to match, for almost all pairs of nearly isometric shapes ($94\%$ of
  pairs in TOSCA dataset and all pairs in SCAPE dataset).
  Note that our method is not limited to bilateral symmetry, as can 
  be seen in the two examples on the left, where similar appendages 
  of the octopuses and ants are matched to each other.
  
  Figure~\ref{fig:gallery_noniso} shows region-based correspondence
  between shapes from different categories which have high isometric
  distortion.  Note that even though the shapes have large variations
  in their intrinsic geometry (and sometimes pose), in most cases the
  segments are cut in similar locations of the shape (e.g. above the
  neck, below the knee, etc), although we stress that our goal is not
  necessarily to extract semantically meaningul segmentations.

Figure~\ref{fig:gallery_pcd} shows region correspondence between triangular meshes 
and point clouds, and between pairs of point clouds. The models vary in body types, 
pose, and sampling density, and include sampling noise. We discuss
these results in more detail in
Section~\ref{sec:pc_evaluation}.

In Figure~\ref{fig:gallery_top} we show examples of matching shapes
with different topological structure, i.e. missing/additional parts
and shapes of different genus.  Our method discards correspondences
with low confidence rather than forcing a match for every region. In
these results, as well as in the results below, parts of shapes that
are not matched are shown in black, e.g. the tail of the armadillo on
the right and the arms of the point cloud on the right.  We argue that
incomplete matching is still useful for certain applications,
including point-to-point map computation, as shown in Section~\ref{sec:p2p}.  Note that in the bottom right pair of point clouds in
Figure~\ref{fig:gallery_top}, both the legs and the arms of the left
shape are touching which drastically changes the shape graph.  Another
example of matching shapes with substantially different shape graphs
is shown in Figure~\ref{fig:symmetry_non_isometric}, along with the
shape graphs for each shape. Note that our method matches similar
sections of the shapes while leaving dissimilar sections unmatched.  }

\begin{figure}[t]
\centering
	\begin{tabular}{@{}c@{\,}c@{}}

		\includegraphics[width=0.45\linewidth]{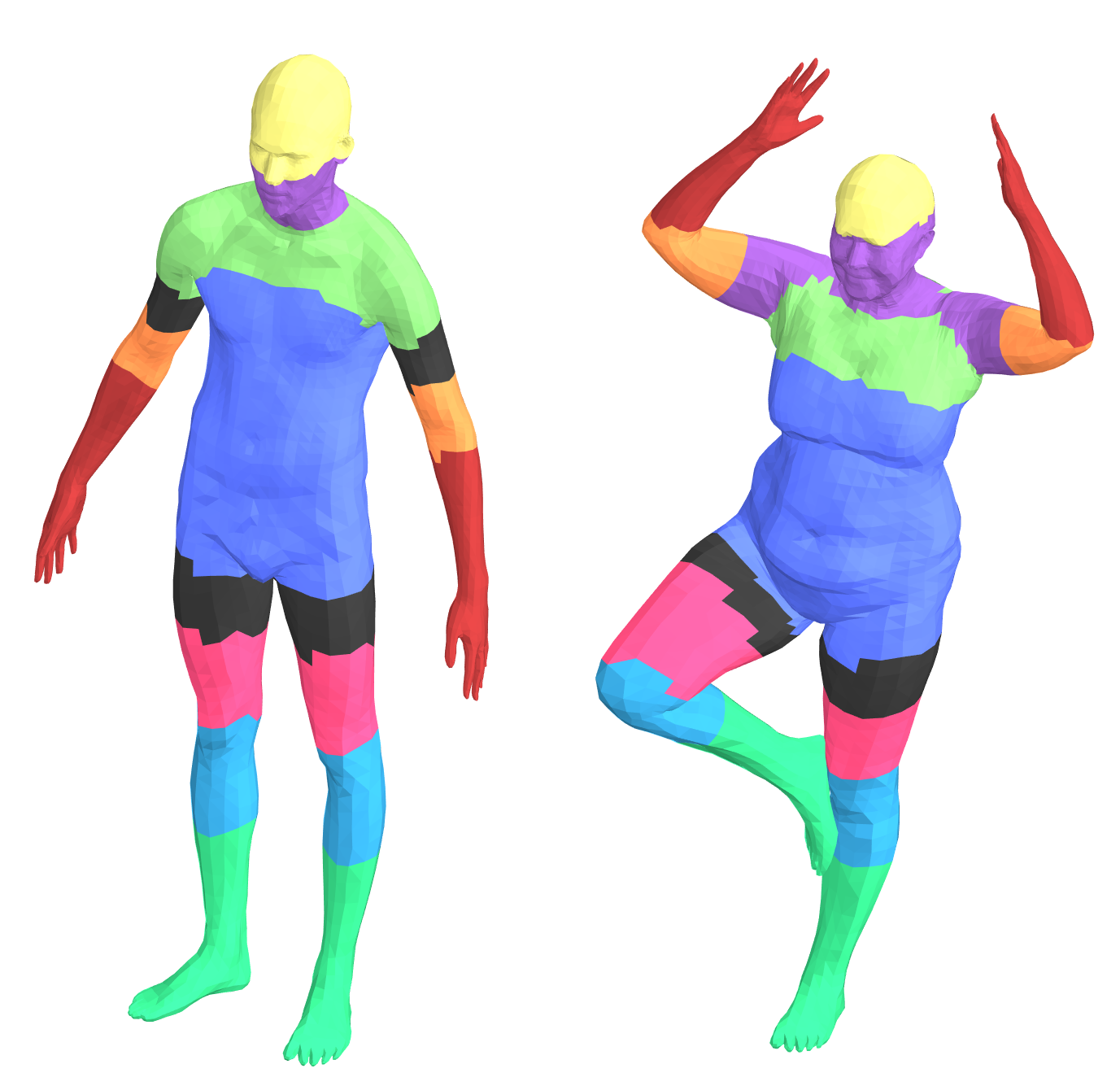} &
		\includegraphics[width=0.45\linewidth]{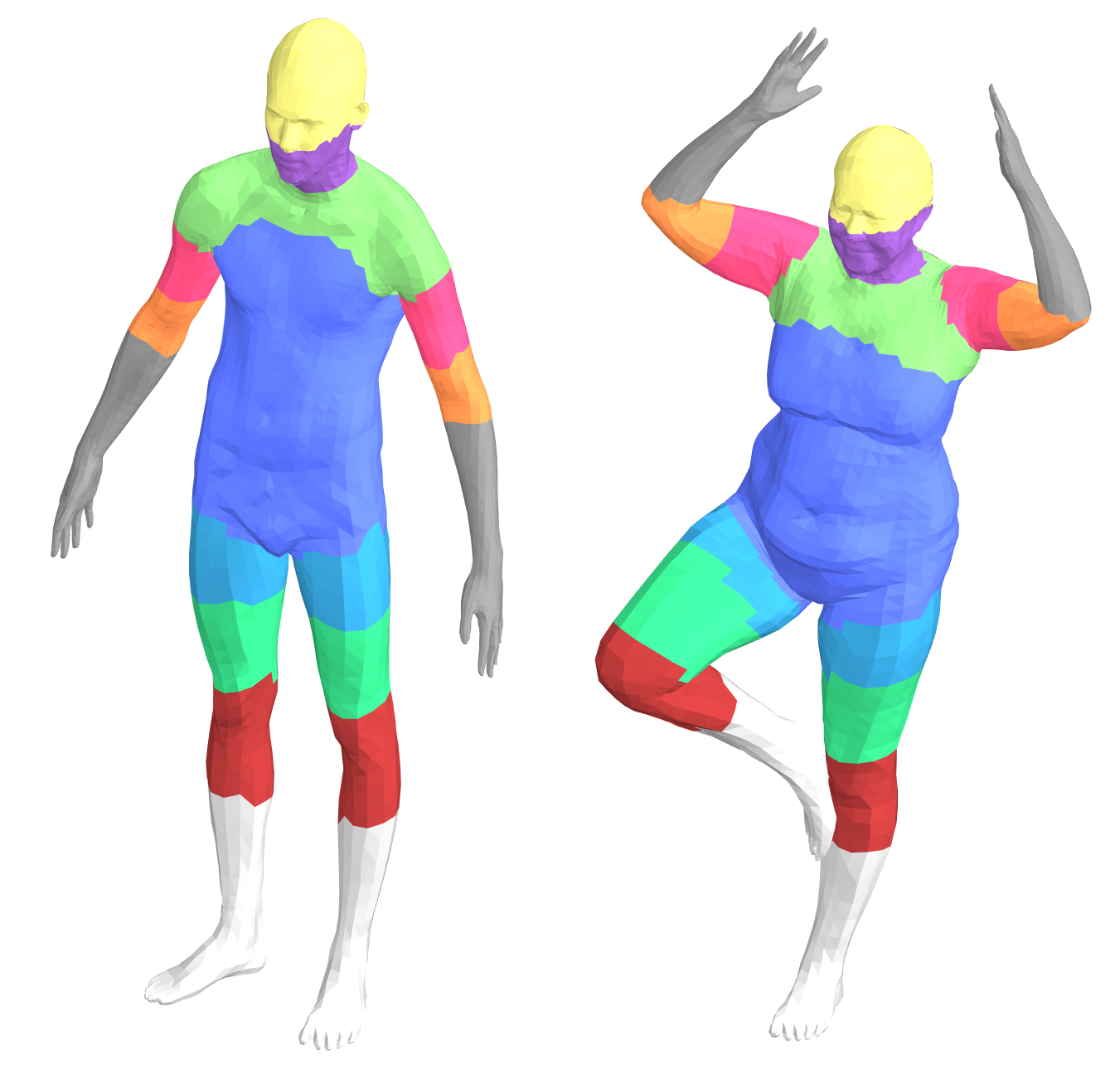} \\

		(a) single time step  & (b) multiple time steps  \\
        (accuracy = 0.7651)   & (accuracy = 0.9079) \\
	\end{tabular}	
\caption[Effect of using multi-dimensional descriptors]{Effect of using multi-dimensional descriptors. (a) a pair of shapes matched using HKS with a single time step. (b) the same shapes matched using a range of time steps.}
\label{fig:multidim}
\vspace{-3mm}
\end{figure}

\paragraph*{Multi-dimensional descriptors.}
In Figure~\ref{fig:multidim} we show the effect of using multi-dimensional descriptors. In (a), the shapes were matched using HKS descriptor with a single time step $t=0.1$. In (b), the shapes were matched using a range of $10$ time steps from $t=0.03$ to $t=0.3$. Note that the location of the cut between segments is more accurate when using a range of time steps. This also results in shape graphs that are more similar to each other and therefore easier to match correctly. We compute the accuracy of the matching using ground truth data (see Section~\ref{sec:region}).

\begin{figure}[t]
\centering
		\includegraphics[width=\linewidth]{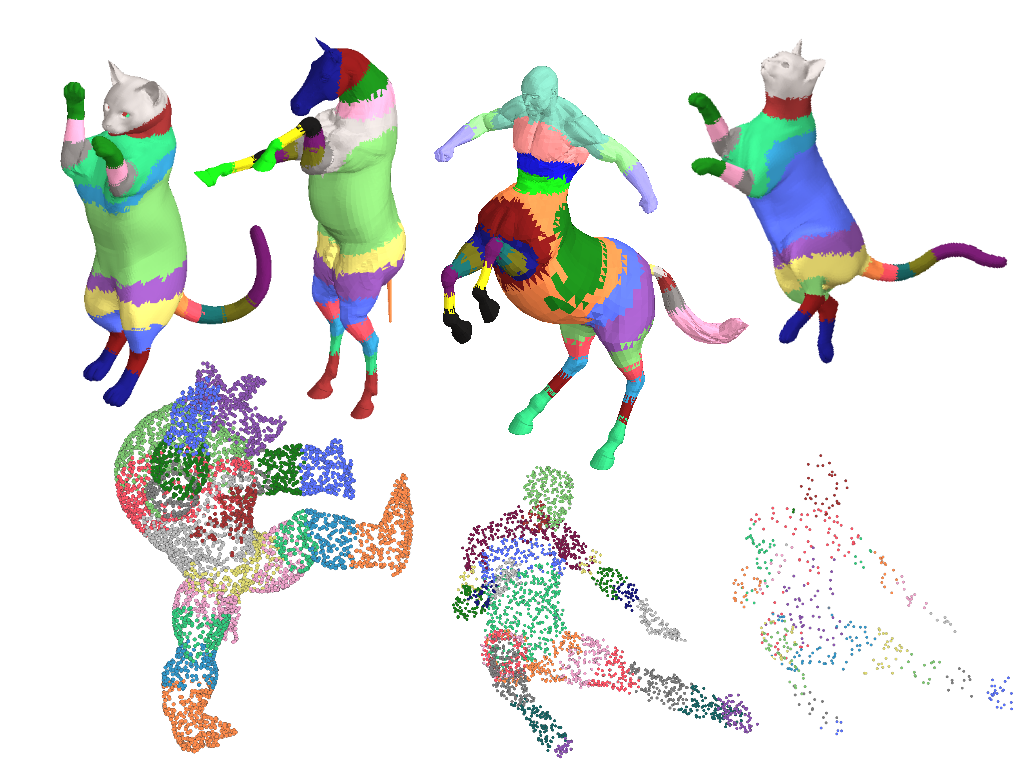}
\caption[Examples of region-level symmetry detection]{Examples of region-level symmetry detection.}
\label{fig:symmetry_detection}
\vspace{-3mm}
\end{figure}

\paragraph*{Symmetry detection.}
In Figure~\ref{fig:symmetry_detection} we show a few examples of
region-wise symmetry detection obtained with our method. In this case
we adopt the same pipeline as described above but use it to match a shape graph to itself, and simply assign the unmatched nodes to themselves. Note that our method is robust to near-isometric distortion. 
In the centaur shape, it can be seen that our method distinguishes well between similar
elements with structural differences such as arms, hind legs and front legs. 
The same method can be used for point clouds; note that the symmetric matching is robust to significantly different sampling densities, as can be seen by the two shapes on the bottom right. 
Finally, note that our method can produce segments which are small enough to indicate symmetry between feature points. 

\begin{table}[t]
\centering
\setlength{\tabcolsep}{1mm}
\begin{tabular}{lccccc} \toprule
  Category		& Ours	& BIM & BIM-FPS & STB & Reeb \\ \midrule
  {\bfseries TOSCA} & & \\
	\; cat 			&	0.8146			& 	\textbf{0.8337}		& 0.7858	& 0.6646	& 0.1101	\\
	\; centaur 		&	0.8719			& 	\textbf{0.9040}		& 0.8400	& 0.7261	& 0.2372	\\
	\; david     	&	\textbf{0.9483}	& 	0.9112				& 0.8600	& 			& 0.0177	\\
	\; dog 			&	\textbf{0.9155}	& 	0.8524				& 0.8118	& 0.5470 	& 0.0920	\\
	\; gorilla 		&	\textbf{0.9178}	& 	0.8966				& 0.8634	& 0.4527	& 0.0166	\\
	\; horse 		&	0.8959			& 	\textbf{0.9030}		& 0.8789	& 0.7463	& 0.1106	\\
	\; michael 		&	\textbf{0.9372}	& 	0.9045				& 0.8614	&			& 0.0552	\\
	\; victoria 	&	\textbf{0.9192}	& 	0.9037				& 0.8668	&			& 0.1025	\\
	\; wolf      	&	0.9570			& 	\textbf{0.9816}		& 0.9723	& 0.9180	& 0.2446	\\ 
    \; \emph{number of segments} & \emph{19.22} &	-	& -		& \emph{9.10}	& \emph{20.68}		\\ \midrule

	{\bfseries SCAPE}		&	\textbf{0.8704}	&	0.8295 		& 0.7279	& 0.8241	& 0.1790	\\
    \; \emph{number of segments} & \emph{17.32} &	-	& -		& \emph{9.95}	& \emph{6.24}		\\ \midrule

	{\bfseries FAUST}	& & \\
	\; same category 		&	\textbf{0.9223}		& 0.8913 	& 0.8785	& 0.7803	& 0.1700	\\
	\; different category 	&	\textbf{0.8473}		& 0.7868 	& 0.6805	& 0.6922	& 0.1213	\\
    \; \; same pose			&   \textbf{0.8975}		& 0.8139 	& 0.8393	& 0.7430	& 0.1545 	\\
    \; \; different pose	&   \textbf{0.8384}		& 0.7820 	& 0.6524	& 0.6836	& 0.1154	\\
    \; \textbf{total}		&	\textbf{0.8814}		& 0.8344 	& 0.7706	& 0.7002	& 0.1434	\\
    \; \emph{number of segments} & \emph{16.88} &	-	& -		& \emph{9.59}	& \emph{25.89}		\\

\bottomrule
\end{tabular} 
\caption{\minorrev{\change{Comparison of our symmetric matching to BIM~\cite{kim11}, stable regions~\cite{ganapathi16},
      and the baseline Reeb graph method (accuracy, higher is better). We also report the average number of segments produced by each method in each
      category, except for BIM, which produces pointwise maps.}}}
 \label{tbl:symmetry_bim}
 \vspace{-3mm}
\end{table}

\begin{figure}[t]
\centering

	\begin{minipage}{0.42\linewidth}
	    \centering
	    \includegraphics[width=\linewidth]{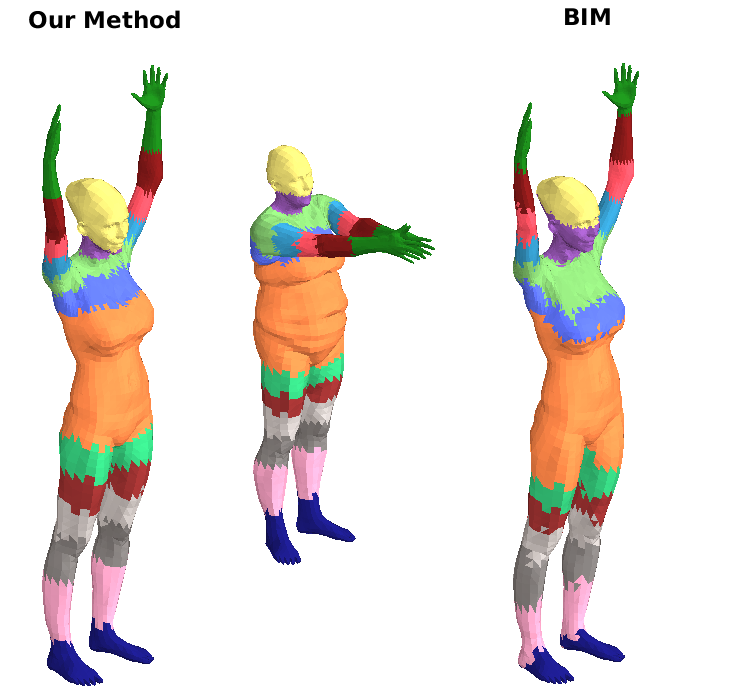}
	    (a)
	\end{minipage}
	\hspace{18pt}
	\begin{minipage}{0.42\linewidth}
	    \centering
	    \includegraphics[width=\linewidth]{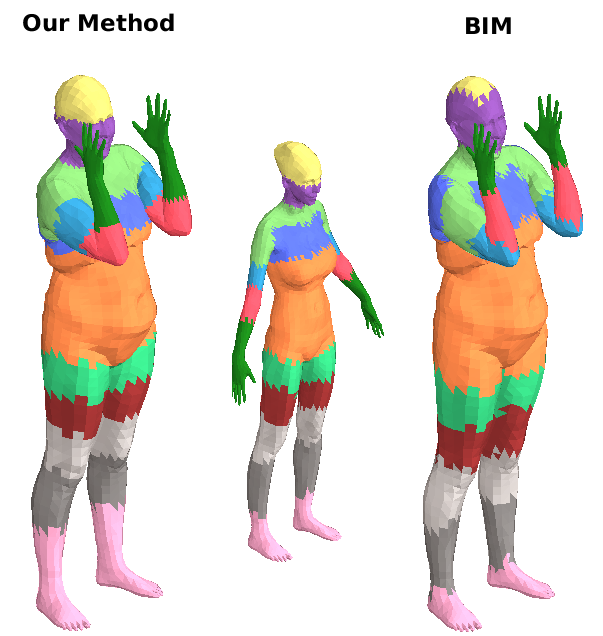}
	    (b)
	\end{minipage}
	\begin{minipage}{0.7\linewidth}
	    \centering
	    \includegraphics[width=\linewidth]{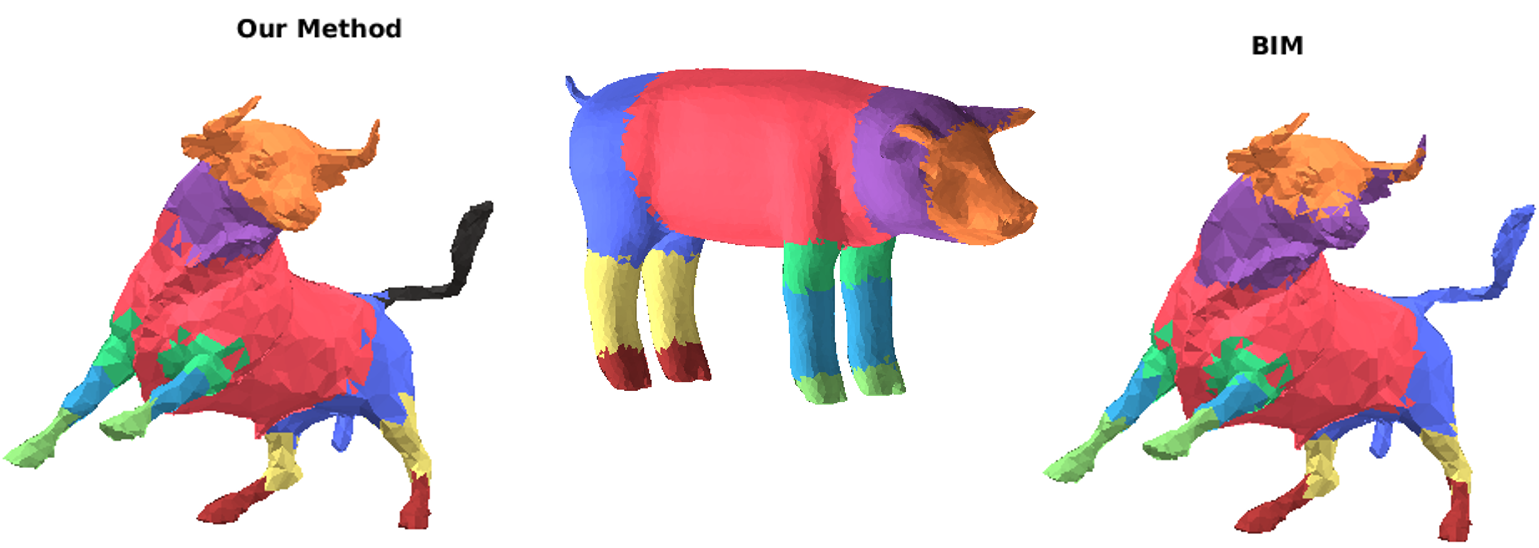}
	    (c)
	\end{minipage}

\caption[Comparison of Symmetry Aware Correspondence to BIM]{Comparison of our method to BIM. In each subfigure, the segments of the central shape were transferred to the shape on the left using our method and to the shape on the right using BIM. Matched segments are shown in the same color.}
\label{fig:symmetry_slipping}
\vspace{-1mm}
\end{figure}

\begin{figure}[t]
\centering

	\begin{tabular}{@{}c@{\,}c@{\,}c@{}}

		\includegraphics[height=3cm]{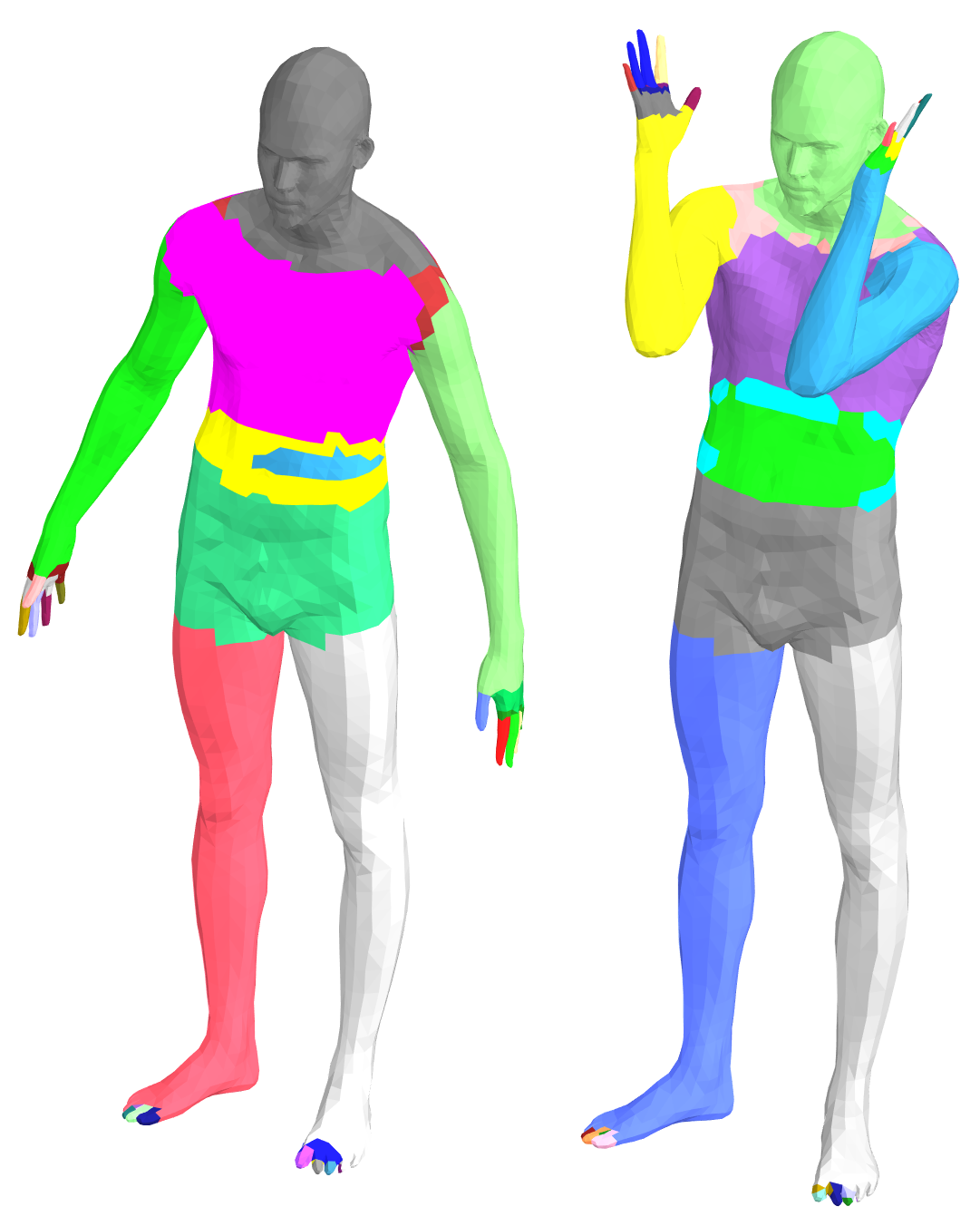} &
		\includegraphics[height=3cm]{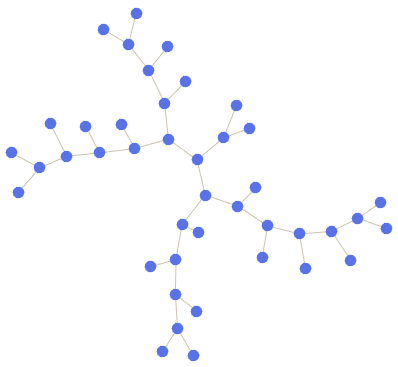} &
		\includegraphics[height=3cm]{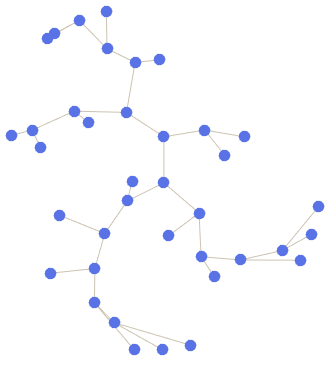} \\

	\end{tabular}	

        \caption[Reeb graphs]{\change{Reeb graphs of two shapes in the same category.}}
\label{fig:reeb}
\vspace{-3mm}

\end{figure}

\subsection{Region Accuracy}
\label{sec:region}
To evaluate the consistency of the regions and the correctness of the matching between them, we use
a ground-truth mapping between vertices and count the number of vertices which are in the same
region in the source and target shape, weighted by the area covered by each vertex. 
\change{
We use the same evaluation for two other baseline methods: the stable region method 
of~\cite{ganapathi16} (STB), and a method based on Reeb graphs,
which are matched using the algorithm described in Section~\ref{sec:matching}. 
The Reeb graphs were computed based on a single HKS descriptor, using code provided by the authors of~\cite{doraiswamy13}. One of the advantages of our method is the use of multidimensional descriptors rather than a scalar function.
Note that~\cite{ganapathi16} typically produces very large regions, which are sometimes discontinuous. 
\minorrev{This is evident by the number of different regions the methods produce on average, which can be seen in Table~\ref{tbl:symmetry_bim}.}
These segments do not compare to ours in their usefulness as they are not discriminative enough. This is not reflected in the evaluation score that does not consider the segment size, thus providing an inherent advantage to large segments.
We were only able to run the stable region method on some of the sets since it is computationally expensive.

We also compare our method with BIM~\cite{kim11}, which does not output segments. For this,
we map the segments our method outputs for the first shape onto the second shape using
the point-to-point map. We count the percent of vertices which are in the correct segment, weighted by the vertex area.}
We also evaluate BIM over a voronoi decomposition of the shape where segment centers are selected 
using farthest point sampling of the shape (BIM-FPS). For this measure we use the same number of segments
that our method produces.

Table~\ref{tbl:symmetry_bim} shows a comparison of the different methods on the TOSCA~\cite{tosca}, SCAPE~\cite{scape} and FAUST~\cite{faust}
datasets.
In most categories, the regions provided by our method are more accurately transferred to the 
target shape. In particular, our method excels in categories where there is a lot of non-isometric 
distortion, such as shapes from different categories and different poses in the FAUST dataset.
A few examples of the difference in region accuracy are shown in Figure~\ref{fig:symmetry_slipping}.
Note that the regions in our method are more consistent with the
source shape, especially in shape extremities such as the head,
arms and legs.
\change{Reeb graphs are often too noisy to be matched correctly. In Figure~\ref{fig:reeb} we show
how shapes within the same category can result in significantly different Reeb graphs even when using a robust feature
function such as the HKS. The regions on the shapes are colored randomly.
}

\subsection{Matching Point Clouds}
\label{sec:pc_evaluation}
Our method works seamlessly over point clouds as well as triangular meshes. \change{There are two main potential
  challenges} that may affect the performance on point clouds compared to triangular meshes. First, there might be
inconsistencies in the descriptor function due to sampling density or noise. Second, \change{the computation of
  connected components can lead to errors in region estimation. Nevertheless, as we show below, our method is robust
  enough to produce accurate results despite these challenges.} For point clouds, we approximate the shape connectivity
by connecting each vertex to its $k$ nearest neighbors (we used $k=6$ in all of our experiments). Therefore, separate
shape parts which are nearby may be considered as a single connected component, thus changing the shape graph.

\begin{figure}[t]
\centering

	\includegraphics[width=0.9\linewidth]{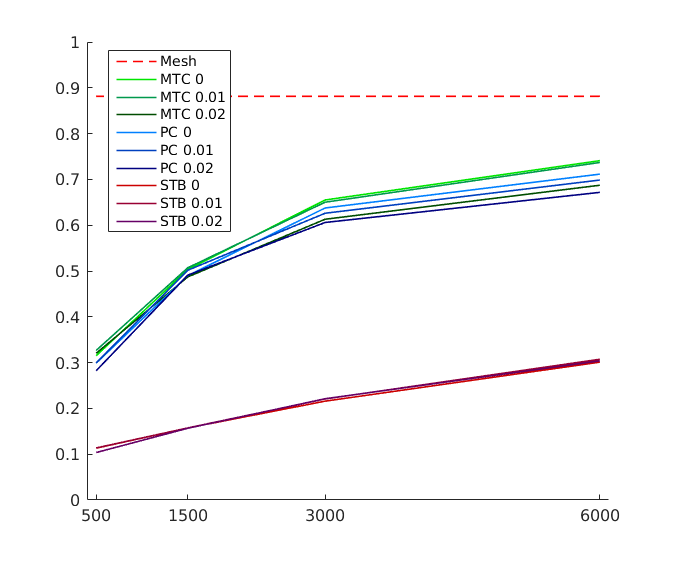}

\caption{
\change{Stability of matching point clouds with varying sampling densities and noise levels, for matching a triangular mesh to a point cloud (MTC), matching two point clouds (PC) and matching two point clouds using the stable region method (STB). The dotted line marked as \emph{Mesh} shows the average accuracy for matching two meshes over all shapes in the dataset. 
\label{fig:pointclouds}\vspace{-3mm}}}

\end{figure}

We perform extensive experiments to evaluate the method's stability under sampling density and
noise. In addition, we compare our method with the stable region method~\cite{ganapathi16} which is
also applicable to point clouds, as it is based on bi-clustering of points. In Figure~\ref{fig:pointclouds} we show the performance of both methods for various noise levels and sampling rates of the shapes in the FAUST dataset. We sample the surface of both shapes uniformly using $6000$, $3000$, $1500$ and $500$ points (we sample uniformly on the surface of the mesh,
including inside the triangles). For each sampling density we apply noise by shifting every vertex by up to $0\%$, $1\%$ or $2\%$ of the shape width in every direction. We repeat each trial four times and show the average of all trials in the graph. We evaluate the results by mapping each point in the point cloud to its nearest vertex in the mesh, and transferring the segmentation from each point cloud to its corresponding mesh. We then compute the accuracy as the percent of vertices that are mapped correctly on the mesh, excluding vertices for which no corresponding points exist on the point cloud. For each noise level, we show the average accuracy as a factor of the sampling density. In this experiment, we used a single HKS descriptor with time step $t=0.1$. The results are presented in the blue lines marked as \texttt{PC}.
We run the same experiments for the stable region method~\cite{ganapathi16}, using the same single HKS descriptor as input. The results are presented in the red lines marked as \texttt{STB}.
We also show results of matching triangle meshes to point clouds. For each pair of shapes, we use the
triangle mesh as the source and the point cloud as the target, sampled with the same parameters and number of trials as above. The results are shown in green lines marked as \texttt{MTC}.

\begin{table}[t]
\centering
\setlength{\tabcolsep}{1.7mm}
\begin{tabular}{lccccc} \toprule

	Category	&  WKS	& 	HKS		& WKS + SEGS  & SEGS	& SYM \\ \midrule
    TOSCA	&	0.0480  & 	0.0613 	&	0.0266			& \textbf{0.0266}	& 0.0391 \\
    SCAPE	&	0.1624	&	0.2569	& 	\textbf{0.0604} & 0.0605			& 0.0958 \\
    FAUST	&	0.0670 	&	0.1184	& 	0.0306 			& \textbf{0.0305}	& 0.0481 \\
	 \bottomrule
\end{tabular} 
\caption[Evaluation of point-to-point maps.]{\change{Evaluation of point-to-point maps computed by functional maps, using WKS, HKS, our corresponding regions with WKS (WKS + SEGS), our regions without WKS descriptors (SEGS), and our symmetric regions (SYM). The numbers represent the average geodesic error (lower is better).  }}
 \label{tbl:p2p}
 \vspace{-2mm}
\end{table}

\subsection{Point-to-point maps}
\label{sec:p2p}
The correspondence between regions can be used to enhance point-to-point maps between the shapes.
To this end, we apply the symmetry breaking heuristic described in Section~\ref{seg:sym_breaking}
and use the functional map framework to generate point-to-point maps.

To evaluate the point-to-point maps, we measure the average geodesic error between points in the
target mesh and their ground truth correspondence. In~\cite{ovsjanikov12}, the functional maps
framework was evaluated using WKS descriptors and functional constraints based on matching segments.
We compute point-to-point maps using WKS descriptors only (without using regions), \change{HKS
  descriptors only (without regions),} using only constraints based on our regions, and using both
WKS descriptors and region-based constraints.  \change{For the HKS descriptors, we provide the same
  descriptors and parameters as input to our region correspondence method and to functional maps.}
The results are presented in Table~\ref{tbl:p2p}. Our results show a significant improvement of the
geodesic error when using our regions. \minorrev{Although not reported, the performance when combining
  HKS and WKS without region correspondences is worse in all shape categories than that of WKS
  alone across a choice of relative weights of the two descriptors.} Interestingly, when applying the region-based constraints, the results do not further
improve with the addition of WKS descriptors. This suggests that our regions are fine and accurate
enough to capture a lot of the information contained in the point-specific WKS descriptors.

\begin{figure}[t]
\centering

	\includegraphics[width=1\linewidth]{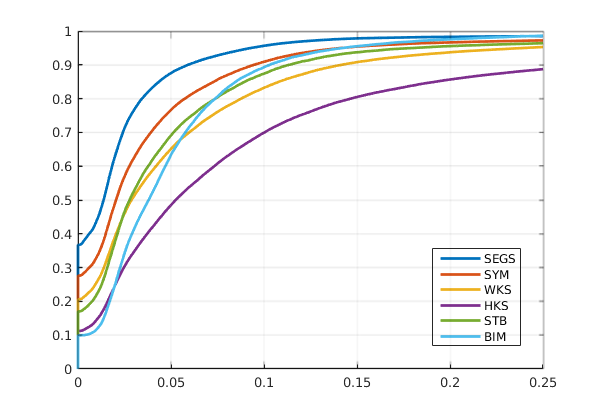}

\vspace{-1mm}
\caption{\change{Average error of the point-to-point maps generated by running the functional maps
    framework~\cite{ovsjanikov12} on our one-to-one correspondence (\texttt{SEGS}), our symmetric correspondence without
    symmetry breaking (\texttt{SYM}), \texttt{WKS} and \texttt{HKS} descriptors, and~\cite{ganapathi16} (\texttt{STB}), and the error of~\cite{kim11} (\texttt{BIM}).
\label{fig:faust_fmaps}
}}
\end{figure}

We also compare the point-to-point correspondences obtained by using 
our region correspondence within the functional maps framework on the
FAUST dataset \cite{faust} and summarize the results in Figure~\ref{fig:faust_fmaps}.
Here, we show the fraction of point-wise matches obtained using different
methods that are within some geodesic distance threshold of the ground truth
(x-axis of the figure).
\change{
We compare the results obtained using our one-to-one region correspondences (\texttt{SEGS}),
our symmetric region correspondences (\texttt{SYM}),
\texttt{WKS} descriptors only, 
\texttt{HKS} descriptors only (with the exact same parameters we use for the region correspondence),
and the region correspondences obtained with the recent stable regions 
approach ~\cite{ganapathi16}, as well as Blended Intrinsic Maps for reference \cite{kim11}. 
We omit the results of using region correspondences without \texttt{WKS} descriptors
from this graph as it merges with the plot of \texttt{SEGS} at this resolution.
Note that our region correspondences provide significantly more information
than the ones produced by stable regions method 
and result in significant improvement in quality of correspondences compared
to other approaches, for both the symmetric region correspondences and one-to-one correspondences.
This is again because the region matches produced by our method are both
accurate and precise, unlike the regions produced by ~\cite{ganapathi16}, which often aggregate
large parts of the shapes, and thus lose their informativeness for pointwise matches.

\paragraph*{Summary of results.}
The quality of the point-to-point map is likely to decrease when erroneous region correspondence is used as input.  The
quality of the maps of FAUST shapes improved in $96\%$ of the cases and decreased in $4\%$, which suggests that the vast
majority of the shapes are matched correctly. On average, the error was reduced by $54\%$, from $0.067$ to $0.0306$.  In
$9.8\%$ of the cases our method resulted in incomplete segment correspondences (i.e., not covering the entire
shape). Nevertheless, even when considering those cases alone (24 shape pairs), the quality improved in $83.3\%$ (20
pairs), reducing the error by $3.8$ times on average. In the 4 remaining cases, the error was increased by $3.3$ times
on average due to incorrect segment matches.}

\begin{table}[t]
\centering
\setlength{\tabcolsep}{1.7mm}
\begin{tabular}{l c c} \toprule
  {\bfseries Method} 			& {\bfseries 7K} 	& {\bfseries 52K} \\ \midrule
	{\bfseries Our method}		& 					& \\
	\quad Symmetric matching  	& 11.35				& 63.11 \\
    \quad Post-processing 		& 4.90				& 99.94 \\
    \quad Total 				& 16.25				& 163.06 \\
    {\bfseries BIM} 			& 28.11 			& 295.86 \\
    {\bfseries Stable regions} 	& 236.93 			& N/A \\

\bottomrule
\end{tabular} 
\caption{\change{Comparison of average running time (in seconds).} 
\vspace{-5mm}}
 \label{tab:timing}
\end{table}

\subsection{Timing}
\label{sec:timing}
In Table~\ref{tab:timing}, we report the average running time of our method, the stable region method~\cite{ganapathi16}, and BIM~\cite{kim11},
\change{for shapes in the FAUST dataset with 7K vertices and a subset of TOSCA shapes with 52K vertices (the stable region methods could not be run for these shapes due to memory limitations).}
All experiments were
run on a machine with Intel Xeon 3GHz processor and 64Gb of RAM. Note that for some
applications only the symmetric matching is necessary, since the post-processing part of our
method is only necessary for producing a one-to-one matching between regions. In comparison, the stable region method only
produces a symmetric matching. Nevertheless, even with this
post-processing our method is significantly faster than both methods.


\section{Conclusion, Limitations  \& Future Work}

We present a method for computing region-level correspondence between shapes. The structure of the shapes is revealed by
their decomposition into regions and used to compute the correspondence at the region level. Our method is both
computationally efficient and robust in the presence of shape changes, such as differences in body types and extreme
pose changes.

Our method outperforms state-of-the-art shape correspondence methods within the domain of matching shape regions.
Moreover, we show that region-level correspondence can be used to improve the quality of point-to-point maps. A notable
advantage of our method is its capability of handling different shape representations, such as matching triangular
meshes to point clouds. We evaluate our method on point clouds with various sampling densities and noise levels, and
show that it is quite robust to noisy sampling.

\change{ Some cases still pose a challenge for our method, as it is designed to find correspondences between shapes of
similar structure. Partial shapes and shapes with different topology (i.e. touching body parts) may cause a drastic
change in the shape graphs such that there is no clear correspondence between the graphs. When the missing or additional
parts are symmetric to an existing part, a correspondence can be found, as seen in Figures~\ref{fig:gallery_top}
and~\ref{fig:symmetry_non_isometric}. However, in more extreme cases when a large portion of the shape is missing, our
approach will fail to provide a meaningful correspondence. As can be seen in Figure~\ref{fig:pointclouds}, as the
quality of the sampled shape deteriorate, the region correspondence becomes less reliable. This is also due to the fact
that noise and lack of information in the point cloud yields different shape graphs which can not be matched correctly.
Another limitation of our approach is that structural data may not be enough to distinguish between parts, even for
isometric shapes. For example, in Figure~\ref{fig:gallery_noniso}, the tail of the giraffe is matched to the hind legs
since their corresponding branches in the shape graph are symmetric and cannot be distinguished. However, as
demonstrated in the quantitative evaluation, this only has a minor effect on the usefulness of our method in practice. }

An interesting direction for future work is to explore more challenging \emph{partial} matching scenarios, and consider
vastly different shape representations, such as 2D sketches vs. 3D point clouds together with our robust shape graph
construction.

\section*{Acknowledgments}
\minorrev{The authors would like thank Prof. Daniel Cohen-Or for useful discussions, as well as Dorian
  Nogneng and Vignesh Ganapathi-Subramanian for help with the experimental evaluation.} Parts of
this work were supported by the Chateaubriand Fellowship, Marie-Curie CIG-334283, a CNRS chaire
d'excellence, chaire Jean Marjoulet from Ecole Polytechnique, FUI project TANDEM 2,  a Google Focused Research Award and the ERC Starting Grant EXPROTEA (StG-2017-758800).


\bibliographystyle{eg-alpha-doi}

\bibliography{references}


\end{document}